\documentclass[12pt]{amsart}
\usepackage[english]{babel}
\usepackage{amssymb}

\numberwithin{equation}{section}

\usepackage{amsmath}
\usepackage{amscd}
\usepackage{amssymb}

\def\C{{\bf C}}
\def\bra{\langle}
\def\ket{\rangle}
\newcommand{\lam}{\lambda}
\newcommand{\up}[1]{{#1}^1,{#1}^2,{#1}^3}
\newcommand{\ammm}{|\Aut(\mu^1)||\Aut(\mu^2)||\Aut(\mu^3)|}
\newcommand{\pa}{\partial}
\newcommand{\cal}{\mathcal}

\newcommand{\bP}{{\Bbb P}}

\newcommand{\bZ}{{\Bbb Z}}

\newcommand{\cF}{{\cal F}}
\newcommand{\cG}{{\cal G}}

\newcommand{\cM}{{\cal M}}
\newcommand{\cO}{{\cal O}}
\newcommand{\cP}{{\cal P}}

\newcommand{\cW}{{\cal W}}
\newcommand{\Mbar}{\overline{\cM}}

\newcommand{\ep}{\epsilon}
\newcommand{\Q}{\mathbb{Q}}
\newcommand{\HG}{\mbox{\it HG}\,}

\DeclareMathOperator{\Aut}{Aut}

\newtheorem{theorem}{Theorem}[section]

\newtheorem{lemma}[theorem]{Lemma}
\newtheorem{coro}[theorem]{Corollary}
\newtheorem{conj}[theorem]{Conjecture}

\theoremstyle{remark}

\theoremstyle{definition}

\renewcommand{\phi}{\varphi}

\def\P{{\bf P}}

\newcommand{\R}{{\mathbb R}}

\begin{document}

\title{Localization and Conjectures from String Duality}
\author{Kefeng Liu}
\address{Center of Mathematical Sciences, Zhejiang University, Hangzhou,
China\\ and
Department of Mathematics\\
University of California at Los Angeles\\ Los Angeles, CA
90095-1555, USA\\
Email: liu@math.ucla.edu, liu@cms.zju.edu.cn}

\date{}

\thanks{Proceedings of the 23rd International Conference of Differential
Geometric Methods in Theoretical Physics Tianjin, 20 - 26 August
2005. The author is supported by the NSF and NSFC}

\maketitle

\numberwithin{equation}{section}


\begin{abstract} We describe the applications of localization methods, in
particular the functorial localization formula, in the proofs of
several conjectures from string theory. Functorial localization
formula pushes the computations on complicated moduli spaces to
simple moduli spaces. It is a key technique in the proof of the
general mirror formulas, the proof of the Hori-Vafa formulas for
explicit expressions of basic hypergeometric series of homogeneous
manifolds, the proof of the Mari\~no-Vafa formula, its
generalizations to two partition analogue. We will also discuss our
development of the mathematical theory of topological vertex and
simple localization proofs of the ELSV formula and Witten
conjecture.
\end{abstract}

\newcommand{\M}{{\mathcal M}}
\section{Introduction}\label{intro}

According to string theorists, String Theory, as the most promising
candidate for the grand unification of all fundamental forces in the
nature, should be the final theory of the world, and should be
unique. But now there are {five} different looking string theories.
As argued by physicists, these theories should be equivalent, in a
way dual to each other. On the other hand all previous theories like
the Yang-Mills and the Chern-Simons theory should be parts of string
theory. In particular their partition functions should be equal or
equivalent to each other in the sense that they are equal after
certain transformation. To compute partition functions, physicists
use localization technique, a modern version of residue theorem, on
infinite dimensional spaces. More precisely they apply localization
formally to path integrals which is not well-defined yet in
mathematics. In many cases such computations reduce the path
integrals to certain integrals of various Chern classes on various
finite dimensional moduli spaces, such as the moduli spaces of
stable maps and the moduli spaces of vector bundles. The
identifications of these partition functions among different
theories have produced many surprisingly beautiful mathematical
formulas like the famous mirror formula \cite{CdGP}, as well as the
Mari\~no-Vafa formula \cite{Mar-Vaf}.

 The mathematical proofs of these conjectural formulas from the string
  duality also depend on localization techniques on these various
   finite dimensional moduli spaces. The purpose of this note is
to discuss our works on the subject. I will briefly discuss the
proof of the mirror conjecture and its generalizations, the proof of
the Hori-Vafa formula, the proof of the Marin\~o-Vafa formula and
its generalizations, the related topological vertex theory
\cite{AKMV} \cite{LLLZ}, and simple localization proofs of the ELSV
formula and the Witten conjecture \cite{Kon}. More precisely we will
use
   localization formulas in various form to compute the integrals
   of Chern classes on moduli
spaces, and to prove those conjectures from string duality. For the
proofs of these conjectures such as the mirror formula, the
Mari\~no-Vafa formula and the theory of topological vertex, we note
that many aspects of mathematics are involved, such as the
Chern-Simons knot invariants, combinatorics of symmetric groups,
representations of Kac-Moody algebras, Calabi-Yau manifolds,
geometry and topology of moduli space of stable maps, etc. The
spirit of our results is the duality among various string theories.
In particular the duality between IIA and IIB string theory gives
the mirror formulas, the duality between gauge theory, Chern-Simons
theory and the Calabi-Yau geometry in string theory leads the
Mari\~no-Vafa conjecture and the theory of topological vertex.

Localization techniques have been very successful in proving many
conjectures from physics, see my ICM 2002 lecture \cite{Liu} for
more examples. One of our major tools in the proofs of these
conjectures is the functorial localization formula which is a
variation of the classical localization formula, it transfers
computations on complicated spaces to simple spaces, and connects
computations of mathematicians and physicists.

In this note we will discuss the following results:
\begin{enumerate}

\item[1.] The proof of the mirror formulas and its generalizations
which we call the mirror principle. The mirror principle implies all
of the conjectural mirror formulas of counting rational curves for
toric manifolds and their Calabi-Yau submanifolds from string
theory. In this case we apply the functorial localization formula to
the map from the nonlinear moduli space to the linearized moduli
space. This transfers the computations of integrals on complicated
moduli space of stable maps to computations on rather simple spaces
like projective spaces. From this the proof of the mirror formula
and its generalizations become conceptually clean and simple.

In fact the functorial localization formula was first found and used
in Lian-Liu-Yau's proof of the mirror conjecture.

\item[2.] The proof of the Hori-Vafa conjecture and its
generalizations for Grassmannian and flag manifolds. This conjecture
predicts an explicit formula for the basic hypergeometric series of
a homogeneous manifold in terms of the basic series of a simpler
manifold such as the product of projective spaces. In this case we
use the functorial localization formula twice to transfer the
computations on the complicated moduli spaces of stable maps to the
computations on quot-schemes. The first is a map from moduli space
of stable maps to product of projective spaces, and another one is a
map from the quot-scheme into the same product of projective spaces.
A key observation we had is that these two maps have the same image.

This approach was first sketched in \cite{LLY3}, the details for
Grassmannians were carried out in \cite{LLLY} and \cite{B-CF-K}. The
most general case of flag manifolds was carried out in \cite{ChLLY2}
and \cite{B-CF-K2}.

\item[3.]The proof of the Mari\~{n}o-Vafa conjecture on Hodge
integrals in \cite{LLZ1}. This conjecture gives a closed formula for
the generating series of a class of triple Hodge integrals for all
genera and any number of marked points in terms of the Chern-Simons
knot invariant of the unknot. This formula was conjectured by M.
Mari\~{n}o and C. Vafa in \cite{Mar-Vaf} based on the duality
between large $N$ Chern-Simons theory and string theory. Many Hodge
integral identities, including the ELSV formula for Hurwitz numbers
\cite{ELSV} and the $\lambda_g$ conjecture \cite{FP2}, can be
obtained by taking various limits of the Mari\~{n}o-Vafa formula
\cite{LLZ2}. The Mari\~{n}o-Vafa formula was first proved by
applying the functorial localization formula to the branch morphism
from the moduli space of relative stable maps to a projective space.

\item[4.] The proof of the generalization of the Mari\~no-Vafa formula
to two partitions cases, and the theory of topological vertex. The
mathematical theory of topological vertex was motivated by the
physical theory as first developed by the Vafa group \cite{AKMV},
who has been working on string duality for the past several years.
Topological vertex theory is a high point of their work starting
from their geometric engineering theory and Witten's conjecture that
Chern-Simons theory is a string theory \cite{Wit}. While the
Marin\~o-Vafa formula gives a close formula for the generating
series of triple Hodge integrals on the moduli spaces of all genera
and any number marked points, topological vertex \cite{LLLZ} gives
the most effective ways to compute the Gromov-Witten invariants of
any open toric Calabi-Yau manifolds. Recently Pan Peng was able to
use our results on topological vertex to give a complete proof of
the Gopakumar-Vafa integrality conjecture for any open toric
Calabi-Yau manifolds \cite{Pan}. Kim also used our technique to
derive new effective recursion formulas for Hodge integrals on the
moduli spaces of stable curves \cite{Kim}.

\item[5.] We describe a very simple proof of the ELSV formula \cite{ELSV}
following our proof of the Mari\~no-Vafa formula, by using the
cut-and-join equation from localization and combinatorics. The proof
of the ELSV formula is particularly easy by using functorial
localization, it is reduced to the fact that the push-forward in
equivariant cohomology of a constant between two equal dimensional
varieties is still constant. We will also show how to directly
derive the ELSV formula from the Mari\~no-Vafa formula by taking a
direct limit.

\item[6.] By using functorial localization formula we have the simple
proofs of the Witten conjecture \cite{Kon}. Our simple proof of the
Witten conjecture in \cite{KL} is to study the asymptotic expansion
of the simple cut-and-join equation for one Hodge integrals which is
derived from functorial localization. This immediately gives a
recursion formula which implies both the Virasoro constraints and
the KdV relation satisfied by the generating series of the $\psi$
integrals.

\end{enumerate}

I will start with brief discussions about the proofs of the mirror
conjecture and the Hori-Vafa formula for Grassmannians, then I will
go to the proofs of the Marin\~o-Vafa conjecture and its
generalizations to two partitions and the topological vertex theory.
After that we discuss the simple proofs of the ELSV formula and the
Witten conjecture. This note is basically a detailed account of my
plenary lecture at the International Conference of Differential
Geometry Method in Theoretical Physics held in August 2005. It is an
much more expanded version of a previous survey I wrote for the 2004
International Complex Geometry Conference held in the Eastern Normal
University of China. I would like to thank the organizers of the
conferences, especially Professor Chunming Bai, Professor Shengli
Tan, Professor Weiping Zhang and Professor Zhijie Chen for their
hospitality during my visits. I would also like to thank my
collaborators for the past 10 years, Bong Lian, Shing-Tung Yau,
Chien-Hao Liu, Melissa C.-C. Liu, Jian Zhou, Jun Li, Yon Seo Kim for
the wonderful experience in solving these conjectures and to develop
the theory together.

\section{Localization} In this section we will explain
 the {\em Functorial Localization Formula}.
We start with a review of the Atiyah-Bott localization formula.
Recall that the definition of equivariant cohomology group for a
manifold $X$ with a torus $T$ action:
$$
H_T^*(X)=H^*(X\times_T ET)
$$
where $ET$ is the universal bundle of $T$, we will use $\R$ or $\Q$
as coefficients through this note.

\medskip

\paragraph{\bf Example}
We know $ES^1= S^{\infty}$. If $S^1$ acts on $\P^n$ by
$$
\lambda\cdot[Z_0,\ldots,Z_n]=
[\lambda^{w_0}Z_0,\ldots,\lambda^{w_n}Z_n],
$$
with $w_0,\cdots, w_n$ as weights, then
$$
H^*_{S^1}(\P^n;\Q)\cong\Q[H,u]/ \bra(H-w_0u)\cdots(H-w_nu)\ket
$$
where $u$ is the generator of $H^*(BS^1, \Q)$. We have the following
important {\em Atiyah-Bott Localization Formula}:

\begin{theorem}

For $\omega\in H^*_T(X)$ an equivariant cohomology class, we have
$$
\omega=\sum_{E} i_{E*}\left(\frac{i_E^*\omega}{e_T(E/X)}\right).
$$
where $E$ runs over all connected components of $T$ fixed points
set, $i_E$ denotes the inclusion map, $i_E^*\ i_{E*}$ denote the
pull-back and push-forward in equivariant cohomology.

\end{theorem}

This formula is very effective in the computations of integrals on
manifolds with torus $T$ symmetry. The idea of localization is
fundamental in many subjects of geometry. In fact Atiyah and Witten
proposed to formally apply this localization formula to loop spaces
and the natural $S^1$-action, from which one gets the Atiyah-Singer
index formula. In fact the Chern characters can be interpreted as
equivariant forms on loop space, and the $\hat{A}$-class is the
inverse of the equivariant Euler class of the normal bundle of $X$
in its loop space $LX$:
$$
e_T(X/LX)^{-1}\sim \hat{A}(X),
$$
which follows from the normalized infinite product formula
$$
\left(\prod_{n\neq 0}(x+n)\right)^{-1}\sim\frac{x}{\sin x}.
$$
I observed in \cite{Liu1} that the normalized product

$$\prod_{m,n}(x+m+n\tau) =2q^{\frac{1}{8}}\sin(\pi x)\\
\cdot \prod_{j=1}^\infty (1-q^j)(1-e^{2\pi ix}q^j)(1-e^{-2\pi
ix}q^j),$$
 where $q=e^{2\pi i\tau}$, also has deep geometric meaning.
 This formula is the Eisenstein formula. It can be viewed as
 a double loop space analogue of the Atiyah-Witten observation.
This formula gives the basic Jacobi $\theta$-function. As observed
by in \cite{Liu1}, formally this gives the $\hat{A}$-class of the
loop space, and the Witten genus which is defined to be the index of
the Dirac operator on the loop space:
$$
e_T(X/LLX)\sim \hat{W}(X),
$$
where $LLX$ is the double loop space, the space of maps from
$S^1\times S^1$ into $X$. $\hat{W}(X)$ is the Witten class. See
\cite{Liu1} for more detail.

The variation of the localization formula we will use in various
situations is the following {\em Functorial Localization Formula}

\begin{theorem}
Let $X$ and $Y$ be two manifolds with torus action. Let $f:\
X\rightarrow Y$ be an equivariant map. Given $F\subset Y$ a fixed
component, let $E\subset f^{-1}(F)$ be those fixed components inside
$f^{-1}(F)$. Let $f_0=f|_E$, then for $\omega\in H_T^*(X)$ an
equivariant cohomology class, we have the following identity on $F$:
$$
{f_0}_*[\frac{i_E^*\omega}{e_T(E/X)}]=\frac{i_F^*(f_*\omega)}{e_T(F/Y)}.
$$
\end{theorem}

This formula will be applied to various settings to prove various
conjectures from physics. It first appeared in \cite{LLY1}. In many
cases we will use a virtual version of this formula. It is used to
push computations on complicated moduli spaces to simpler moduli
spaces. A $K$-theory version of the functorial localization formula
also holds \cite{LLY2}, interesting applications are expected.

\medskip
\paragraph{\bf Remark}
Consider the diagram:
$$
   \begin{array}{ccc}
    H^*_T(X) & \stackrel{f_*}{\longrightarrow}
       & H^*_T(Y) \\[.6ex]
     \downarrow\,\mbox{\scriptsize ${i_E}^*$}
       & & \downarrow\,\mbox{\scriptsize ${i_F}^*$} \\
    H^*_T(E)
       & \stackrel{{f_0}_*}{\longrightarrow}
       & H^*_T(F)\,.
   \end{array}
$$
The functorial localization formula is like Riemann-Roch with the
inverted equivariant Euler classes of the normal bundle as
"weights", in a way similar to the Todd class for the Riemann-Roch
formula. In fact if we formally apply this formula to the map
between the loop spaces of $X$ and $Y$, equivariant with respect to
the rotation of the circle, we do formally get the differentiable
Riemann-Roch formula. We believe this can be done rigorously by
following Bismut's proof of the index formula which made rigorous of
the above argument of Atiyah-Witten.

\section{ Mirror Principle }

There have been many discussions of mirror principle in the
literature. Here we only give a brief account of the main ideas of
the setup and proof of the mirror principle. We will use two most
interesting examples to illustrate the algorithm. These two examples
give proofs of the mirror formulas for toric manifolds as
conjectured by string theorists.

The goal of mirror principle is to compute the characteristic
numbers on moduli spaces of stable maps in terms of certain
hypergeometric type series. This was motivated by mirror symmetry in
string theory. The most interesting case is the counting of the
numbers of curves which corresponds to the computations of Euler
numbers. More generally we would like to compute the characteristic
numbers and classes induced from the general Hirzebruch
multiplicative classes such as the total Chern classes. The
computations of integrals on moduli spaces of those classes pulled
back through evaluation maps at the marked points and the general
Gromov-Witten invariants can also be considered as part of mirror
principle. Our hope is to develop a "black-box" method which makes
easy the computations of the characteristic numbers and the
Gromov-Witten invariants.

The general set-up of mirror principle is as follows. Let $X$ be a
projective manifold, $\overline{{\cal M}}_{g, k}(d, X)$ be the
moduli space of stable maps of genus $g$ and degree $d$ with $k$
marked points into $X$, modulo the obvious equivalence. The points
in $\overline{\cM}_{g,k}(d, X)$ are triples $ (f; C;\, x_1, \cdots,
x_k)$ where $ f: C\rightarrow X$ is a degree $d$ holomorphic map and
$x_1, \cdots, x_k$ are $k$ distinct smooth points on the genus $g$
curve $C$. The homology class $f_*([C])=d\in H_2(X, \mathbb{Z})$ is
identified as integral index $d=(d_1, \cdots, d_n)$ by choosing a
basis of $H_2(X,\bZ)$, dual to the K\"ahler classes.

In general the moduli space may be very singular, and may even have
different dimension for different components. To define integrals on
such singular spaces, we need the virtual fundamental cycle of
Li-Tian \cite{LT}, and also Behrend-Fantechi \cite{B-F} which we
denote by $[\overline{\cM}_{g, k}(d,X)]^v$. This is a homology class
of the expected dimension
$$
2\left(c_1(TX)[d]+(\dim_{\mathbb{C}} X-3)(1-g)+k\right)
$$
on $\overline{\cM}_{g, k}(d,X)$.

Let us consider the case $k=0$ first. Note that the expected
dimension of the virtual fundamental cycle is $0$ if $X$ is a
Calabi-Yau 3-fold. This is the most interesting case for string
theory.

The starting data of mirror principle are as follows. Let $V$ be a
concavex bundle on $X$ which we defined as the direct sum of a
positive and a negative bundle on $X$. Then $V$ induces a sequence
of vector bundles $V^g_d$ on $\overline{{\cal M}}_{g, 0}(d, X)$
whose fiber at $(f; C; x_1, \cdots, x_k)$ is given by $H^0(C,
f^*V)\oplus H^1(C, f^*V)$. Let $b$ be a multiplicative
characteristic class. So far for all applications in string theory,
$b$ is the Euler class.

The problem of mirror principle is to compute
$$
K^g_d= \int_{[\overline{\cM}_{g, 0}(d, X)]^v} b(V^g_d).
$$
More precisely we want to compute the generating series
$$
F(T, \lambda)=\sum_{d,\, g} K^g_d \,\lambda^g ~e^{d\cdot T}
$$
in terms of certain hypergeometric type series. Here $\lambda$,
$T=(T_1,\cdots, ,T_n)$ are formal variables.

The most famous formula in the subject is the Candelas formula as
conjectured by P. Candelas, X. de la Ossa, P. Green, and L. Parkes
\cite{CdGP}. This formula changed the history of the subject. More
precisely, Candelas formula considers the genus 0 curves, that is,
we want to compute the so-called {\em A-model potential} of a
Calabi-Yau 3-fold $M$ given by
$$
\cF_0(T)=\sum_{d\in H_2(M;\bZ)}K^0_d \, e^{d\cdot T},
$$
where $T=(T_1,\ldots, T_n)$ are considered as the coordinates of the
Kahler moduli of $M$, and $K^0_d$ is the genus zero, degree $d$
invariant of $M$ which gives the numbers of rational curves of all
degree through the multiple cover formula \cite{LLY1}. The famous
mirror conjecture asserts that there exists a mirror Calabi-Yau
3-fold $M'$ with {\em B-model potential} $\cG(T)$, which can be
computed by period integrals,  such that
$$
\cF(T)=\cG(t),
$$
where $t$ accounts for coordinates of complex moduli of $M'$. The
map $t\mapsto T$ is called the {\em mirror map}. In the toric case,
the period integrals are explicit solutions to the GKZ-system, that
is the Gelfand-Kapranov-Zelevinsky hypergeometric series. While the
A-series are usually very difficult to compute, the B-series are
very easy to get. This is the magic of the mirror formula. We will
discuss the proof of the mirror principle which includes the proof
of the mirror formula.

The key ingredients for the proof of the mirror principle consists
of
\begin{enumerate}
\item  Linear and non-linear moduli spaces;
\item  Euler data and hypergeometric (HG) Euler data.
\end{enumerate}

More precisely, the non-linear moduli is the moduli space $M_d^g(X)$
which is the stable map moduli of degree $(1, d)$ and genus $g$ into
$\mathbf{P}^1\times X$. A point in $M_d^g(X)$ consists of a pair
$(f, C): \, f: C\rightarrow \mathbf{P}^1\times X$ with $C$ a genus
$g$ (nodal) curve, modulo obvious equivalence. The linearized moduli
$W_d$ for toric $X$ were first introduced by Witten and used by
Aspinwall-Morrison to do approximating computations.

\medskip

\paragraph{\bf Example} Consider the projective space $\mathbf{P}^n$ with
homogeneous coordinate $[z_0, \cdots, z_n]$. Then the linearized
moduli $W_d$ is defined as projective space with coordinates
$$[f_0(w_0, w_1), \cdots, f_n(w_0, w_1)]$$ where $f_j(w_0, w_1)$'s are
homogeneous polynomials of degree $d$.

This is the simplest compactification of the moduli spaces of degree
$d$ maps from $\mathbf{P}^1$ into $\mathbf{P}^n$. The following
lemma is important. See \cite{LLY4} for its proof. The $g=0$ case
was given in \cite{Gi} and in \cite{LLY1}.

\begin{lemma}
There exists an explicit equivariant collapsing map
$$
\varphi:\  M^g_d(\mathbf{P}^n)\longrightarrow W_d.
$$
\end{lemma}

For general projective manifold $X$, the nonlinear moduli $M_d^g(X)$
can be embedded into $M^g_d(\mathbf{P}^n)$. The nonlinear moduli
$M_d^g(X)$ is very "singular" and complicated, but the linear moduli
$W_d$ is smooth and simple. The embedding induces a map of
$M^g_d(X)$ to $W_d$. Functorial localization formula pushed the
computations onto $W_d$. Usually mathematical computations should be
done on the moduli of stable maps, while physicists tried to use the
linearized moduli to approximate the computations. So functorial
localization formula connects the computations of mathematicians and
physicists. In some sense the mirror symmetry formula is more or
less the comparison of computations on nonlinear and linearized
moduli.

Mirror principle has been proved to hold for balloon manifolds. A
projective manifold $X$ is called balloon manifold if it admits a
torus action with isolated fixed points, and if the following
conditions hold. Let
$$
H=(H_1, \cdots, H_k)
$$
be a basis of equivariant Kahler classes such that
\begin{enumerate}
\item the restrictions $H(p)\neq H(q)$ for any two fixed points $p\neq q$;
\item the tangent bundle $T_pX$ has linearly independent weights for any
fixed point $p$.
\end{enumerate}
This notion was introduced by Goresky-Kottwitz-MacPherson.

\begin{theorem}
Mirror principle holds for balloon manifolds and for any concavex
bundles.
\end{theorem}

\paragraph{\bf Remarks}
\begin{enumerate}
\item[1.] All toric manifolds are balloon manifolds. For
$g=0$ we can identify the hypergeometric series explicitly. Higher
genus cases need more work to identify such series.
\item[2.] For toric manifolds and $g=0$, mirror principle implies all of
the mirror conjectural formulas from string theory.
\item [3.] For Grassmannian manifolds, the explicit mirror formula is
given by the Hori-Vafa formula to be discussed in Section
\ref{sec:hori-vafa}.
\item [4.] The case of direct sum of positive line bundles on
$\mathbf{P}^n$, including the Candelas formula. For proofs of this
formula from two different points of view, see \cite{LLY1} and
\cite{Gi, P,BDPP}; for a comparison of the two points of view, see
\cite{CK}.
\end{enumerate}

Now we briefly discuss the proof of the mirror principle. The main
idea is to apply the functorial localization formula to $\varphi$,
the collapsing map and the pull-back class $\omega=\pi^*b(V_d^g)$,
where $\pi: \ \ {{ M}}^g_d(X) \rightarrow
  \overline{{\cal M}}_{g,0}(d, X)$ is the natural projection.

Such classes satisfy certain induction property. To be precise we
introduce the notion of {\em Euler Data}, which naturally appears on
the right hand side of the functorial localization formula,
$Q_d=\varphi_!(\pi^*b(V_d^g))$ which is a sequence of polynomials in
equivariant cohomology rings of the linearized moduli spaces with
simple quadratic relations. We also considered their restrictions to
$X$.

From functorial localization formula we prove that, by knowing
 the Euler data $Q_d$ we can determine the $K_d^g$. On the other hand,
 there is another much simpler Euler data, the
{\em HG Euler data} $P_d$, which coincides with $Q_d$ on the
"generic" part of the nonlinear moduli. We prove that the quadratic
relations and the coincidence
 on generic part determine the Euler data uniquely up to certain degree.
 We also know that $Q_d$ always have the right degree for $g=0$.
We then use mirror transformation to reduce the degrees of the HG
Euler data $P_d$. From these we deduced the mirror principle.

\medskip

\paragraph{\bf Remarks}
\begin{enumerate}
\item[1.] Both the denominator and the numerator in the HG
series, the generating series of the HG Euler data, are equivariant
Euler classes. Especially the denominator is exactly from the
localization formula. This is easily seen from the functorial
localization formula.
\item[2.] The quadratic relation of Euler data, which naturally comes
from gluing and functorial localization on the A-model side, is
closely related to special geometry, and is similar to the
Bershadsky-Cecotti-Ooguri-Vafa's {\em holomorphic anomaly} equation
on the B-model side. Such relation can determine the
polynomial Euler data up to certain degree.\\
It is an interesting task to use special geometry to understand the
mirror principle computations, especially the mirror transformation
as a coordinate change.
\item[3.] The Mari\~{n}o-Vafa formula to be discussed later is needed
 to determine the hypergeometric Euler data for higher
genus computations in mirror principle. The Mari\~{n}o-Vafa formula
comes from the duality between Chern-Simons theory and Gromov-Witten
theory. This duality and the matrix model for Chern-Simons theory
indicate that mirror principle may have matrix model description.
\end{enumerate}

Let us use two examples to illustrate the algorithm of mirror
principle.

\medskip

\paragraph{\bf Example} Consider the Calabi-Yau quintic in $\mathbf{P}^4$.
In this case
$$
P_d=\prod_{m=0}^{5d}(5\kappa -m\alpha)
$$
with $\alpha$ can be considered as the weight of the $S^1$ action on
$\mathbf{P}^1$, and $\kappa$ denotes the generator of the
equivariant cohomology ring of $W_d$.

The starting data of the mirror principle in this case is
$V=\mathcal{O}(5)$ on $X={\mathbf P}^4$. The hypergeometric series,
after taking $\alpha =-1$, is given by
$$
HG[B](t)= e^{H\,t}\sum^\infty_{d=0}
\frac{\prod_{m=0}^{5d}(5H+m)}{\prod_{m=1}^{d}(H+m)^5}\, e^{d\,t},
$$
where $H$ is the hyperplane class on $\mathbf{P}^4$ and $t$ is a
formal parameter.

We introduce the series
$$
{\cal{F}}(T) =\frac{5}{6}T^3 +\sum_{d>0} K^0_d\, e^{d\,T}.
$$
The algorithm is as follows. Take the expansion in $H$:
$$
HG[B](t)= H\{ f_0(t) +f_1(t)H +f_2(t) H^2 +f_3(t)H^3\},
$$
from which we have the famous {\em Candelas Formula}: With
$T=f_1/f_0$,
$$
\cF(T)=\frac{5}{2}(\frac{f_1}{f_0}\frac{f_2}{f_0}-\frac{f_3}{f_0}).
$$

\medskip

\paragraph{\bf Example} Let $X$ be a toric manifold and $g=0$. Let
$D_1,..,D_N$ be the $T$-invariant divisors in $X$. The starting data
consist of $V=\oplus_i L_i$ with $c_1(L_i)\geq0$ and
$c_1(X)=c_1(V)$. Let us take $b(V)=e(V)$ the Euler class. We want to
compute the A-series
$$
A(T)=\sum K^0_d \,e^{d\cdot T}.
$$
The HG Euler series which is the generating series of the HG Euler
data can be easily written down as
$$
B(t)=e^{-H\cdot t}\sum_d \prod_i\prod_{k=0}^{\bra c_1(L_i),d\ket}
(c_1(L_i)-k) {\prod_{\bra D_a,d\ket<0} \prod_{k=0}^{-\bra
D_a,d\ket-1}(D_a+k)\over \prod_{\bra D_a,d\ket\geq0}
\prod_{k=1}^{\bra D_a,d\ket}(D_a-k)}~ e^{d\cdot t}.
$$

Then mirror principle implies that there are explicitly computable
functions $f(t),g(t)$, which define the mirror map, such that
$$
\int_X\left(e^f B(t)-e^{-H\cdot T}e(V)\right) =2A(T)-\sum T_i
{\partial A(T)\over\partial T_i}
$$
where $T=t+g(t)$. From this equation we can easily solve for $A(T)$.

In general we want to compute:
$$
K^g_{d, k}=\int_{[{\cal{M}}_{g, k}(d,
X)]^v}\prod^k_{j=1}ev_j^*\omega_j \,\cdot  b(V^g_d)$$ where
$\omega_j\in H^*(X)$ and $ev_j$ denotes the evaluation map at the
$j$-th marked point. We form a generating series with $t$, $\lambda$
and $\nu$ formal variables,
$$
F(t, \lambda, \nu) =\sum_{d, g,k}K^g_{d,k}e^{dt}\lambda^{2g}\nu^k.
$$
The ultimate mirror principle we want to prove is to compute this
series in terms of certain explicit HG series. It is easy to show
that those classes in the integrand can still be combined to induce
Euler data. Actually the Euler data really encode the geometric
structure of the stable map moduli.

\medskip

We only use one example to illustrate the higher genus mirror
principle.

\medskip

\paragraph{\bf Example} Consider open toric Calabi-Yau manifold, say ${\cal
O}(-3)\rightarrow \mathbf{P}^2$. Here $V={\cal O}(-3)$. Let
$$
Q_d=\sum_{g\geq 0} \varphi_!(\pi^*e_T(V^g_d))\, \lambda^{2g}.
$$
Then it can be shown that the corresponding HG Euler data is given
explicitly by
$$
P_d\, J(\kappa, \alpha, \lambda)J(\kappa-d\alpha, -\alpha, \lambda),
$$
where $P_d$ is exactly the genus $0$ HG Euler data and $J$ is
generating series of Hodge integrals with summation over all genera.
$J$ may be considered as the degree $0$ Euler data. In fact we may
say that the computations of Euler data include computations of all
Gromov-Witten invariants, and even more. Zhou has obtained some
closed formulas. We have proved that the mirror principle holds in
such general setting. The remaining task is to determine the
explicit HG Euler data. But the recently developed topological
vertex theory has given complete closed formulas in terms of the
Chern-Simons invariants. See the discussion in Section 7 for
details.

\medskip

Finally we mention some recent works. First we have constructed
refined linearized moduli space for higher
 genus, the {\em A-twisted moduli stack} $\mathcal{AM}_g(X)$ of genus
 $g$ curves associated to a smooth toric variety $X$, induced from
 the gauged linear sigma model studied by  Witten.

This new moduli space is constructed as follows. A morphism from a
curve of genus $g$ into $X$ corresponds to an equivalence class of
triples $(L_{\rho},u_{\rho},c_m)_{\rho,m}$, where each $L_\rho$ is a
line bundle pulled back from $X$, $u_\rho$ is a section of $L_\rho$
satisfying a non-degeneracy condition, and the collection
$\{c_m\}_m$ gives conditions to compare the sections $u_\rho$ in
different line bundles $L_\rho$, $\mathcal{AM}_g(X)$ is the moduli
space of such data. It is an Artin stack, fibered over the moduli
space of quasi-stable curves \cite{ChLLY1}. We hope to use this
refined moduli to do computations for higher genus mirror principle.

On the other hand, motivated by recent progresses in open string
theory, we are also trying to develop open mirror principle. Open
string theory predicts formulas for the counting of holomorphic
discs with boundary inside a Lagrangian submanifold, more generally
of the counting of the numbers of open Riemann surfaces with
boundary in Lagrangian submanifold. Linearized moduli space for such
data is being constructed which gives a new compactification of such
moduli spaces.

\section{Hori-Vafa Formula}\label{sec:hori-vafa}

In \cite{H-V}, Hori and Vafa generalize the world-sheet aspects of
mirror symmetry to being the equivalence of $d=2$, $N=(2,2)$
supersymmetric field theories (i.e. without imposing the conformal
invariance on the theory). This leads them to a much broader
encompassing picture of
 mirror symmetry. Putting this in the frame
work of abelian gauged linear sigma models (GLSM) of Witten enables
them to link many $d=2$ field theories together. Generalization of
this setting to nonabelian GLSM
 leads them to the following conjecture,
 when the physical path integrals are interpreted appropriately
 mathematically:

\begin{conj}[{Hori-Vafa \cite{H-V}}]
{\em  The hypergeometric series for a given homogeneous space
  $($e.g.\ a Grassmannian manifold$)$ can be reproduced from
  the hypergeometric series of simpler homogeneous spaces
  $($e.g.\ product of projective spaces$)$.
 Similarly for the twisted hypergeometric series that are related
  to the submanifolds in homogeneous spaces.}
\end{conj}

 In other words, different homogeneous spaces (or some simple
quotients of them) can give rise to generalized mirror pairs. A main
object to be understood in the above conjecture is the fundamental
hypergeometric series $\HG[{\mathbf 1}]^X(t)$ associated to the flag
 manifold $X$. Recall that in the computations of mirror principle, the existence
of linearized moduli made easy the computations for toric manifolds.

 An outline of how this series may be computed was given
 in \cite{LLY3} via an extended mirror principle diagram.
To make clear the main ideas we will only focus on the case of
Grassmannian manifolds in this article. The main problem for the
computation is that there is no known good linearized moduli for
Grassmannian or general flag manifolds. To overcome the difficulty
we use the Grothendieck quot scheme to play the role of the
linearized moduli. The method gives a complete proof of the
Hori-Vafa formula in the Grassmannian case.

Let $ev:\ {\overline{{\cal M}}}_{0, 1}(d, X)\rightarrow X$ be the
evaluation map on the moduli space of stable maps with one marked
point, and $c$ the first Chern class of the tangent line at the
marked point. The fundamental hypergeometric series for mirror
formula is given by the push-forward:
$$
ev_*[\frac{1}{\alpha(\alpha-c)}]\in H^*(X)
$$
or more precisely the generating series
$$
HG[1]^X(t) =e^{-tH/\alpha}\sum_{d=0}^\infty
 ev_*[\frac{1}{\alpha(\alpha-c)}]\, e^{dt}.
$$

Assume the linearized moduli exists. Then functorial localization
formula applied to the collapsing map: $\varphi: \ M_d \rightarrow
N_d$, immediately gives the expression as the denominator of the
hypergeometric series.

\medskip

\paragraph{\bf Example} $X= \mathbf{P}^n$, then we have $\varphi_*(1)=1$,
functorial localization immediately gives us
$$
ev_*[\frac{1}{\alpha(\alpha-c)}]=\frac{1}{\prod^d_{m=1}(x-m\alpha)^{n+1}}
$$
where the denominators of both sides are equivariant Euler classes
of normal bundles of the fixed points. Here $x$ denotes the
hyperplane class.

\medskip

For $X=\mathrm{Gr}(k, n)$ or general flag manifolds, no explicit
linearized moduli is known. Hori-Vafa conjectured a formula for
$HG[1]^X(t)$ by which we can compute this series in terms of those
of projective spaces which is the {\em Hori-Vafa formula for
Grassmannians:}

\begin{theorem}
We have
$$
 HG[1]^{\mathrm{Gr}(k, n)}(t)
= \frac{e^{(k-1)\pi\sqrt{-1}\sigma/\alpha} }{\prod_{i<j}(x_i-x_j)}
\cdot\prod_{i<j}\left. (\alpha\frac{\partial}{\partial
x_i}-\alpha\frac{\partial}{\partial
x_j})\right|_{t_i=t+(k-1)\pi\sqrt{-1}}HG[1]^{\mathbf{P}}(t_1,
\cdots,t_k)
$$
where $\mathbf{P}=\mathbf{P}^{n-1}\times \cdots \times
\mathbf{P}^{n-1}$ is product of $k$ copies of the projective spaces,
$\sigma$ is the generator of the divisor classes on $\mathrm{Gr}(k,
n)$ and $x_i$ the hyperplane class of the $i$-th copy
$\mathbf{P}^{n-1}$:
$$
HG[1]^{\mathbf{P}}(t_1,
\cdots,t_k)=\prod^k_{i=1}HG[1]^{\mathbf{P^{n-1}}}(t_i).
$$

\end{theorem}

Now we describe the ideas of the proof of the above formula. As
mentioned above we use another smooth moduli space, the Grothendieck
quot-scheme $Q_d$ to play the role of the linearized moduli, and
apply the functorial localization formula. Here is the general
set-up:

To start, note that the Pl\"{u}cker embedding  $\tau:\;
\mathrm{Gr}(k, n) \rightarrow \mathbf{P}^N$ induces an embedding of
the nonlinear moduli $M_d$ of $\mathrm{Gr}(k, n)$ into that of
$\mathbf{P}^N$. Composite of this map with the collapsing map gives
us a map $\varphi:\; M_d \rightarrow W_d$ into the linearized moduli
space $W_d$ of $\mathbf{P}^N$. On the other hand the Pl\"{u}cker
embedding also induces a map $\psi:\; Q_d \rightarrow W_d.$ We have
the following three crucial lemmas proved in \cite{LLLY}.

\begin{lemma}
The above two maps have the same image in $W_d$: $\mathrm{Im}\, \psi
=\mathrm{Im}\, \varphi$. And all the maps are equivariant with
respect to the induced circle action from $\mathbf{P}^1$.
\end{lemma}

Just as in the mirror principle computations, our next step is to
analyze the fixed points of the circle action induced from
$\mathbf{P}^1$. In particular we need the distinguished fixed point
set to get the equivariant Euler class of its normal bundle. The
distinguished fixed point set in $M_d$ is $\overline{{\cal M}}_{0,
1}(d, \mathrm{Gr}(k, n))$ with equivariant Euler class of its normal
bundle given by $\alpha(\alpha -c)$, and  we know that $\varphi$ is
restricted to $ev$.

\begin{lemma}
The distinguished fixed point set in $Q_d$ is a union: $\cup_s
E_{0s}$, where each $E_{0s}$ is a fiber bundle over $\mathrm{Gr}(k,
n)$ with fiber given by flag manifold.
\end{lemma}

It is a complicated work to determine the fixed point sets $E_{0s}$
and the weights of the circle action on their normal bundles. The
situation for flag manifold cases are much more involved. See
\cite{LLLY} and \cite{ChLLY2} for details.

Now let $p$ denote the projection from $E_{0s}$ onto $\mathrm{Gr}(k,
n)$. Functorial localization formula, applied to $\varphi$ and
$\psi$, gives us the following

\begin{lemma}
We have the equality on $\mathrm{Gr}(k, N)$:
$$
ev_*[\frac{1}{\alpha(\alpha-c)}] =\sum_s
p_*[\frac{1}{e_T({E_{0s}}/Q_d)}]$$ where $e_T(E_{0s}/Q_d)$ is the
equivariant Euler class of the normal bundle of $E_{0s}$ in $Q_d$.
\end{lemma}

Finally we compute $p_*[\frac{1}{e_T(E_{0s}/Q_d)}]$. There are two
different approaches, the first one is by direct computations in
\cite{LLLY}, and another one is by using the well-known Euler
sequences for universal sheaves \cite{B-CF-K}. The second method has
the advantage of being more explicit. Note that
$$
e_T(TQ|_{E_{0s}} -TE_{0s})= e_T(TQ|_{E_{0s}})/e_T(TE_{0s}).
$$
Both $e_T(TQ|_{E_{0s}})$ and $e_T(TE_{0s})$ can be written down
explicitly in terms of the universal bundles on the flag bundle
$E_{0s}= Fl(m_1, \cdots, m_k, S)$ over $\mathrm{Gr}(r, n)$. Here $S$
is the universal bundle on the Grassmannian.

The push-forward by $p$ from $Fl(m_1, \cdots, m_k, S)$ to
$\mathrm{Gr}(r, n)$ is done by an analogue of family localization
formula of Atiyah-Bott, which is given by a sum over the Weyl groups
along the fiber which labels the fixed point sets.

In any case the final formula of degree $d$ is given by
$$
p_*[\frac{1}{e_T(E_{0s}/Q_d)}]=(-1)^{(r-1)d}\sum_{\stackrel{(d_1,\dots
,d_r)}{d_1+...+d_r=d}} \frac{\prod_{1\leq i<j\leq
r}(x_i-x_j+(d_i-d_j)\alpha)} {\prod_{1\leq i<j\leq
r}(x_i-x_j)\prod_{i=1}^r\prod_{l=1}^{d_i} (x_i+l\alpha)^n}.
$$
Here $x_1,...x_r$ are the Chern roots of $S^*$. As a corollary of
our approach, we have the following:

\begin{coro}
 The Hori-Vafa conjecture holds for Grassmannian manifolds.
\end{coro}
See \cite{B-CF-K} and \cite{LLLY} for the details. For the explicit
forms of Hori-Vafa conjecture for general flag manifolds, see
\cite{ChLLY2} and \cite{B-CF-K2}.

\section{ The Mari\~{n}o-Vafa Conjecture}

Our original motivation to study Hodge integrals was to find a
general mirror formula for counting higher genus curves in
Calabi-Yau manifolds. To generalize mirror principle to count the
number of higher genus curves, we need to first compute Hodge
integrals, i.e. the intersection numbers of the $\lambda$ classes
and $\psi$ classes on the Deligne-Mumford moduli space of stable
curves $\Mbar_{g,h}$. This moduli space is possibly the most famous
and most interesting orbifold. It has been studied since Riemann,
and by many Fields medalists for the past 50 years, from many
different point of views. Still many interesting and challenging
problems about the geometry and topology of these moduli spaces
remain unsolved. String theory has motivated many fantastic
conjectures about these moduli spaces including the famous Witten
conjecture which is about the generating series of the integrals of
the $\psi$-classes. We start with the introduction of some
notations.

Recall that a point in $\Mbar_{g,h}$ consists of
$(C,x_1,\ldots,x_h)$, a (nodal) curve $C$ of genus $g$, and $n$
distinguished smooth points on $C$. The Hodge bundle $\mathbb{E}$ is
a rank $g$ vector bundle over $\Mbar_{g,h}$ whose fiber over
$[(C,x_1,\ldots,x_h)]$ is $H^0(C,\omega_C)$, the complex vector
space of holomorphic one forms on $C$. The $\lambda$ classes are the
Chern Classes of $\mathbb{E}$,
$$
\lambda_i=c_i(\mathbb{E})\in H^{2i}(\Mbar_{g,h};\mathbb{Q}).
$$

On the other hand, the cotangent line $T_{x_i}^* C$ of $C$ at the
$i$-th marked point $x_i$ induces a line bundle $\mathbb{L}_i$ over
$\Mbar_{g,h}$. The $\psi$ classes are the Chern classes:
$$
\psi_i=c_1(\mathbb{L}_i)\in H^2(\Mbar_{g,h};\mathbb{Q}).
$$
Introduce the total Chern class
$$
\Lambda_g^\vee(u)=u^g -\lambda_1 u^{g-1}+\cdots+(-1)^g \lambda_g.
$$

The Mari\~{n}o-Vafa formula is about the generating series of the
triple Hodge integrals
$$
\int_{\Mbar_{g,h}}\frac{\Lambda_g^\vee(1)\Lambda_g^\vee(\tau)
\Lambda_g^\vee(-\tau-1)}{\prod_{i=1}^h(1-\mu_i\psi_i)},
$$
where $\tau$ is considered as a parameter here. Later we will see
that it actually comes from the weight of the group action, and also
from the framing of the knot. Taking Taylor expansions in $\tau$ or
in $\mu_i$ one can obtain information on the integrals of the Hodge
classes and the $\psi$-classes. The Marin\~o-Vafa conjecture asserts
that the generating series of such triple Hodge integrals for all
genera and any numbers of marked points can be expressed by a close
formula which is a {\em finite} expression in terms of
representations of symmetric groups, or Chern-Simons knot
invariants.

We remark that the moduli spaces of stable curves have been the
sources of many interests from mathematics to physics. Mumford has
computed some low genus numbers. The Witten conjecture, proved by
Kontsevich \cite{Kon}, is about the integrals of the $\psi$-classes.

Let us briefly recall the background of the conjecture. Mari\~no and
Vafa \cite{Mar-Vaf} made this conjecture based on the large $N$
duality between Chern-Simons and string theory. It starts from the
conifold transition. We consider the resolution of singularity of
the conifold $X$ defined by
\[
\left\{ \left(\begin{array}{cc}x & y \\z & w\end{array}\right)\in
\C^4: xw-yz=0 \right\}
\]
in two different ways:

 (1). {Deformed conifold} $T^* S^3$
\[
\left\{ \left(\begin{array}{cc}x & y \\z & w\end{array}\right)\in
\C^4: xw-yz=\ep \right\}
\]
where $\ep$ a real positive number. This is a symplectic resolution
of the singularity.

(2). {Resolved conifold} by blowing up the singularity, which gives
the total space

$$\tilde{X}=\cO(-1)\oplus\cO(-1)\rightarrow\mathbf{P}^1$$ which is explicitly
given by
\[
\left\{([Z_0,Z_1], \left(\begin{array}{cc}x & y \\z &
w\end{array}\right)) \in \mathbf{P}^1\times\C^4:
\begin{array}{ll}
(x,y)\in [Z_0,Z_1]\\
(z,w)\in [Z_0,Z_1]\end{array}\right\}
\]

\[
\begin{array}{ccc}
\tilde{X} &\subset&\mathbf{P}^1\times\C^4\\
\downarrow & & \downarrow\\
X &\subset & \C^4
\end{array}
\]

The brief history of the development of the conjecture is as
follows. In 1992 Witten first conjectured that the open topological
string theory on the deformed conifold $T^* S^3$ is equivalent to
the Chern-Simons gauge theory on $S^3$. Such idea was pursued
further by Gopakumar and Vafa in 1998, and then by Ooguri and Vafa
in 2000. Based on the above conifold transition, they conjectured
that the open topological string theory on the deformed conifold
$T^*S^3$ is equivalent to the closed topological string theory on
the resolved conifold $\tilde{X}$. Ooguri-Vafa only considered the
zero framing case. Later Marin\~o-Vafa generalized the idea to the
non-zero framing case and discovered the beautiful formula for the
generating series of the triple Hodge integrals. Recently Vafa and
his collaborators systematically developed the theory, and for the
past several years, they developed these duality ideas into the most
effective tool to compute Gromov-Witten invariants on toric
Calabi-Yau manifolds. The high point of their work is the theory of
topological vertex. We refer to \cite{Mar-Vaf} and \cite{AKMV} for
the details of the physical theory and the history of the
development.

Starting with the proof of the Marin\~o-Vafa conjecture \cite{LLZ1},
\cite{LLZ2}, we have developed a rather complete mathematical theory
of topological vertex \cite{LLLZ}. Many interesting consequences
have been derived for the past year. Now let us see how the string
theorists derived mathematical consequence from the above naive idea
of string duality. First the Chern-Simons partition function has the
form

$$ \langle Z(U,V)\rangle = \exp (- F(\lambda, t, V) )$$ where $U$ is
the holonomy of the $U(N)$ Chern-Simons gauge field around the knot
$K\subset S^3$, and  $V$ is an extra $U(M)$ matrix. The partition
function $\langle Z(U,V)\rangle$ gives the Chern-Simons knot
invariants of $K$.

String duality asserts that the function $F(\lambda, t, V)$ should
give the generating series of the open Gromov-Witten invariants of
$(\tilde{X},L_K)$, where $L_K$ is a Lagrangian submanifold of the
resolved conifold $\tilde{X}$ canonically associated to the knot
$K$.  More precisely by applying the t'Hooft large $N$ expansion,
and the "canonical" identifications of parameters similar to mirror
formula, which at level $k$ are given by

$$\lambda =\frac{2\pi}{k+N},\ \ t=\frac{2\pi iN}{k+N},$$
we get the partition function of the topological string theory on
conifold $\tilde{X}$, and then on $\mathbf{P}^1$. which is just the
generating series of the Gromov-Witten invariants. This change of
variables is very striking from the point of view of mathematics.

The special case when $K$ is the unknot is already very interesting.
In non-zero framing it gives the Mari\~no-Vafa conjectural formula.
In this case $\bra Z(U,V)\ket$ was first computed in the zero
framing by Ooguri-Vafa and in any framing $\tau\in\mathbb{Z}$ by
Mari\~{n}o-Vafa \cite{Mar-Vaf}. Comparing with Katz-Liu's
computations of $F(\lambda,t,V)$, Mari\~{n}o-Vafa conjectured the
striking formula about the generating series of the triple Hodge
integrals for all genera and any number of marked points in terms of
the Chern-Simons invariants, or equivalently in terms of the
representations and combinatorics of symmetric groups. It is
interesting to note that the framing in the Mari\~{n}o-Vafa's
computations corresponds to the choice of lifting of the circle
action on the pair $(\tilde{X},L_{\textup{unknot}})$ in Katz-Liu's
localization computations. Both choices are parametrized by an
integer $\tau$ which will be considered as a parameter in the triple
Hodge integrals. Later we will take derivatives with respect to this
parameter to get the cut-and-join equation.

It is natural to ask what mathematical consequence we can have for
general duality, that is for general knots in general three
manifolds, a first naive question is what kind of general Calabi-Yau
manifolds will appear in the duality, in place of the conifold. Some
special cases corresponding to the Seifert manifolds are known by
gluing several copies of conifolds.

Now we give the precise statement of the  Mari\~no-Vafa conjecture,
which is an identity between the geometry of the moduli spaces of
stable curves and Chern-Simons knot invariants, or the combinatorics
of the representation theory of symmetric groups.

Let us first introduce the geometric side. For every partition $\mu
= (\mu_1 \geq \cdots \mu_{l(\mu)} \geq 0)$, we define the triple
Hodge integral to be,

$$ G_{g, \mu}(\tau) = A(\tau)\cdot \int_{\Mbar_{g, l(\mu)}}
\frac{\Lambda^{\vee}_g(1)\Lambda^{\vee}_g(-\tau-1)\Lambda_g^{\vee}(\tau)}
{\prod_{i=1}^{l(\mu)}(1- \mu_i \psi_i)},$$ where the coefficient

 $$A(\tau)= - \frac{\sqrt{-1}^{|\mu|+l(\mu)}}{|\Aut(\mu)|}
 [\tau(\tau+1)]^{l(\mu)-1}\prod_{i=1}^{l(\mu)}\frac{ \prod_{a=1}^{\mu_i-1} (\mu_i
\tau+a)}{(\mu_i-1)!}.$$ The expressions, although very complicated,
arise naturally from localization computations on the moduli spaces
of relative stable maps into $\mathbf{P}^1$ with ramification type
$\mu$ at $\infty$.

We now introduce the generating series

$$ G_{\mu}(\lambda; \tau) = \sum_{g \geq 0}
\lambda^{2g-2+l(\mu)}G_{g, \mu}(\tau).$$ The special case when $g=0$
is given by

$$ \int_{\Mbar_{0, l(\mu)}}
\frac{\Lambda^\vee_0(1)\Lambda^\vee_0(-\tau-1)\Lambda_0^\vee(\tau) }
{\prod_{i=1}^{l(\mu)} (1 -\mu_i \psi_i) } =\int_{\Mbar_{0, l(\mu)}}
\frac{1}{\prod_{i=1}^{l(\mu)}(1 - \mu_i\psi_i)}$$ which is known to
be equal to $|\mu|^{l(\mu)-3} $ for $l(\mu)\geq 3$, and we use this
expression to extend the definition to the case $l(\mu)<3$.

Introduce formal variables $p=(p_1,p_2,\ldots,p_n,\ldots)$, and
define
$$
p_\mu=p_{\mu_1}\cdots p_{\mu_{l(\mu)} }
$$
for any partition $\mu$. These $p_{\mu_j}$ correspond to
$\mathrm{Tr}\, V^{\mu_j}$ in the notations of string theorists. The
generating series for all genera and all possible marked points are
defined to be

$$G(\lambda; \tau; p)  =  \sum_{|\mu| \geq 1}
G_{\mu}(\lambda;\tau)p_{\mu},$$ which encode complete information of
the triple Hodge integrals we are interested in.

Next we introduce the representation theoretical side. Let
$\chi_{\mu}$ denote the character of the irreducible representation
of the symmetric group $S_{|\mu|}$, indexed by $\mu$ with
$|\mu|=\sum_j \mu_j$. Let $C(\mu)$ denote the conjugacy class of
$S_{|\mu|}$ indexed by $\mu$. Introduce

$$ \cW_{\mu}(\lambda) =  \prod_{1 \leq a < b \leq l(\mu)}
\frac{\sin \left[(\mu_a - \mu_b + b - a)\lambda/2\right]}{\sin
\left[(b-a)\lambda/2\right]}\cdot\frac{1}{\prod_{i=1}^{l(\nu)}\prod_{v=1}^{\mu_i}
2 \sin \left[(v-i+l(\mu))\lambda/2\right]}.$$ This has an
interpretation in terms of quantum dimension in Chern-Simons knot
theory.

We define the following generating series

$$R(\lambda; \tau; p) = \sum_{n \geq 1} \frac{(-1)^{n-1}}{n}
\sum_{\mu}[\sum_{\cup_{i=1}^n \mu^i = \mu} \prod_{i=1}^n
\sum_{|\nu^i|=|\mu^i|}
\frac{\chi_{\nu^i}(C(\mu^i))}{z_{\mu^i}}e^{\sqrt{-1}(\tau+\frac{1}{2})\kappa_{\nu^i}\lambda/2}
\cW_{\nu^i}(\lambda) ]p_\mu$$ where $\mu^i$ are sub-partitions of
$\mu$, $z_{\mu} = \prod_j \mu_j!j^{\mu_j}$ and

$$\kappa_{\mu} = |\mu| + \sum_i (\mu_i^2 - 2i\mu_i)$$ for a
partition $\mu$ which is also standard for representation theory of
symmetric groups. There is the relation $z_\mu
=|\Aut(\mu)|\mu_1\cdots\mu_{l(\mu)}$.

Finally we can give the precise statement of {\em the Mari\~no-Vafa
conjecture:}

\begin{conj}
 {\em We have the identity}

$${G(\lambda;\tau;p) = R(\lambda;\tau;p)}.$$

\end{conj}

Before discussing the proof of this conjecture, we first give
several remarks.

\medskip

\paragraph{\bf Remarks:}
\begin{enumerate}

\item[1.] This conjecture is a formula: $\mathrm{ G: Geometry = R:
Representations}, $  and the representations of symmetric groups are
essentially combinatorics.

\item[2.] We note that each $G_\mu(\lambda, \tau)$ is given by a finite
and closed expression in terms of the representations of symmetric
groups:

$$G_\mu(\lambda, \tau)=\sum_{n \geq 1} \frac{(-1)^{n-1}}{n}
\sum_{\cup_{i=1}^n \mu^i = \mu} \prod_{i=1}^n\sum_{|\nu^i|=|\mu^i|}
\frac{\chi_{\nu^i}(C(\mu^i))}{z_{\mu^i}}e^{\sqrt{-1}(\tau+\frac{1}{2})\kappa_{\nu^i}\lambda/2}
\cW_{\nu^i}(\lambda). $$ The generating series $G_\mu(\lambda,
\tau)$ gives the values of the triple Hodge integrals for moduli
spaces of curves of all genera with $l(\mu)$ marked points.

\item[3.] Note that an equivalent expression of this formula is
 the following non-connected generating series. In this situation we have a relatively simpler
combinatorial expression:

$${G}(\lambda;\tau; p)^{\bullet} = {\mathrm{exp}}\,{[G(\lambda;
\tau; p)]}=\sum_{|\mu| \geq 0}[ \sum_{|\nu|=|\mu|}
\frac{\chi_{\nu}(C(\mu))}{z_{\mu}}e^{\sqrt{-1}(\tau+\frac{1}{2})\kappa_{\nu}\lambda/2}
\cW_{\nu}(\lambda)]p_\mu.$$ According to Mari\~no and Vafa, this
formula gives values for all Hodge integrals up to three Hodge
classes. Lu proved that this is right if we combine with some
previously known simple formulas about Hodge integrals.

\item[4.] By taking Taylor expansion in $\tau$ on both sides of the
Mari\~{n}o-Vafa formula, we have derived various Hodge integral
identities in \cite{LLZ3}.

For examples, as easy consequences of the Mari\~no-Vafa formula and
the cut-and-join equation as satisfied by the above generating
series, we have unified simple proofs of the $\lambda_g$ conjecture
by comparing the coefficients in $\tau$ in the Taylor expansions of
the two expressions,

\begin{eqnarray*}
\int_{\Mbar_{g, n}} \psi_1^{k_1} \cdots \psi_n^{k_n}\lambda_g =
\begin{pmatrix} 2g+n-3 \\ k_1, \dots, k_n\end{pmatrix}
\frac{2^{2g-1}- 1}{2^{2g-1}} \frac{|B_{2g}|}{(2g)!},
\end{eqnarray*}
for $k_1 + \cdots + k_n = 2g-3+n$, and the following identities for
Hodge integrals:

$$ \int_{\Mbar_{g}}\lambda_{g-1}^3=\int_{\Mbar_{g}}
\lambda_{g-2}\lambda_{g-1}\lambda_g = \frac{1}{2(2g-2)!}
\frac{|B_{2g-2}|}{2g-2} \frac{|B_{2g}|}{2g}, $$ where $B_{2g}$ are
Bernoulli numbers. And

$$\int_{\Mbar_{g, 1}} \frac{\lambda_{g-1}}{1-\psi_1} = b_g
\sum_{i=1}^{2g-1} \frac{1}{i} - \frac{1}{2}
\sum_{\substack{g_1+g_2=g\\g_1,g_2 >
0}}\frac{(2g_1-1)!(2g_2-1)!}{(2g-1)!} b_{g_1}b_{g_2},$$ where
$$b_g = \begin{cases} 1, & g = 0, \\
 \frac{2^{2g-1}- 1}{2^{2g-1}} \frac{|B_{2g}|}{(2g)!}, & g > 0.
 \end{cases}$$

\end{enumerate}

Now let us look at how we proved this conjecture. This is joint work
with Chiu-Chu Liu, Jian Zhou, see \cite{LLZ} and \cite{LLZ1} for
details.

The {first} proof of this formula is based on the {\em Cut-and-Join}
equation which is a beautiful match of combinatorics and geometry.
The details of the proof is given in \cite{LLZ} and \cite{LLZ1}.
First we look at the combinatorial side. Denote by $[s_1,\cdots,
s_k]$ a $k$-cycle in the permutation group. We have the following
two obvious operations:
\begin{enumerate}

\item[1.] {\em Cut}: a $k$-cycle is cut into an $i$-cycle and a $j$-cycle:

$$[s, t]\cdot [s, s_2,\cdots, s_i, t, t_2, \cdots t_j]=[s, s_2,\cdots, s_i][t, t_2, \cdots
t_j].$$

\item[2.]{\em Join}: an $i$-cycle and a $j$-cycle are joined to an
$(i+j)$-cycle:

$$[s, t]\cdot [s, s_2,\cdots, s_i][t, t_2, \cdots t_j]=[s, s_2,\cdots, s_i, t, t_2, \cdots
t_j].$$

\end{enumerate}
Such operations can be organized into differential equations which
we call the cut-and-join equation.

Now we look at the geometry side. In the moduli spaces of stable
maps, cut and join have the following geometric meaning:

\begin{enumerate}

\item[1.] {\em Cut}: one curve splits into two lower degree or lower
genus curves.

\item[2.] {\em Join}: two curves are joined together to give a higher
genus or higher degree curve.

\end{enumerate}
The combinatorics and geometry of cut-and-join are reflected in the
following two differential equations, which look like heat equation.
It is easy to show that such equation is equivalent to a series of
systems of linear ordinary differential equations by comparing the
coefficients on $p_\mu$. These equations are proved either by easy
and direct computations in combinatorics or by localizations on
moduli spaces of relative stable maps in geometry. In combinatorics,
the proof is given by direct computations and was explored in
combinatorics in the mid 80s and later by Zhou \cite{LLZ} for this
case. The differential operator on the right hand side corresponds
to the cut-and-join operations which we also simply denote by
$(CJ)$.

\begin{lemma}
$$
\frac{\pa R}{\pa
\tau}=\frac{1}{2}\sqrt{-1}\lambda\sum_{i,j=1}^\infty
((i+j)p_ip_j\frac{\pa R}{\pa p_{i+j} } +ij p_{i+j}( \frac{\pa R}{\pa
p_i}\frac{\pa R}{\pa p_j} +\frac{\pa^2 R}{\pa p_i\pa p_j})).
$$
\end{lemma}

On the geometry side the proof of such equation is given by
localization on the moduli spaces of relative stable maps into the
the projective line $\mathbf{P}^1$ with fixed ramifications at
$\infty$:

\begin{lemma}
 $$
\frac{\pa G}{\pa \tau} = \frac{1}{2}
\sqrt{-1}\lambda\sum_{i,j=1}^\infty ((i+j)p_ip_j\frac{\pa G}{\pa
p_{i+j} }+ij p_{i+j}(\frac{\pa G}{\pa p_i}\frac{\pa G}{\pa p_j}
+\frac{\pa^2 G}{\pa p_i\pa p_j})).$$
\end{lemma}
The proof of the above equation is given in \cite{LLZ}. Together with the following\\

\paragraph{ {\bf Initial Value}}: $\tau =0$,
\[
G(\lambda,0,p)=
\sum_{d=1}^\infty\frac{p_d}{2d\sin\left(\frac{\lambda d}{2}\right) }
=R(\lambda,0, p).
\]
which is precisely the Ooguri-Vafa formula and which has been proved
previously for example in \cite{Zho1}, we therefore obtain the
equality which is the Mari\~no-Vafa conjecture by the uniqueness of
the solution:

\begin{theorem}
{We have the identity}

$$G(\lambda;\tau;p)=R(\lambda;\tau;p).$$
\end{theorem}

 During the proof we note that the cut-and-join equation is
encoded in the geometry of the moduli spaces of stable maps. In fact
we later find the convolution formula of the following form, which
is a relation for the disconnected version $G^\bullet
={\mbox{exp}}\,G$,

$$G^\bullet_\mu(\lambda,\tau)=\sum_{|\nu|=|\mu|}\Phi^\bullet_{\mu,\nu}(-\sqrt{-1}\tau\lambda)z_\nu
K^\bullet_\nu(\lambda)$$ where $\Phi^\bullet_{\mu,\nu}$ is the
generating series of double Hurwitz numbers, and $z_\nu$ is the
combinatorial constant appeared in the previous formulas.
Equivalently this gives the explicit solution of the cut-and-join
differential equation with initial value $K^\bullet(\lambda)$, which
is
 the generating series of the integrals of certain Euler classes on the moduli spaces of
relative stable maps to $\mathbf{P}^1$. See \cite{CLiu} for the
derivation of this formula, and see \cite{LLZ2} for the two
partition analogue.

The Witten conjecture as proved by Kontsevich states that the
generating series of the $\psi$-class integrals satisfy infinite
number of differential equations. The remarkable feature of
Mari\~{n}o-Vafa formula is that it gives a finite close formula. In
fact by taking limits in $\tau$ and $\mu_i$'s one can obtain the
Witten conjecture. A much simpler direct proof of the Witten
conjecture was obtained recently by Kim and myself in \cite{KL}. We
directly derived the recursion formula which implies both the
Virasoro relations and the KdV equations.

The same argument as our proof of the conjecture gives a simple and
geometric proof of the ELSV formula for Hurwitz numbers. It reduces
to the fact that the push-forward of $1$ is a constant in
equivariant cohomology for a generically finite-to-one map. See
\cite{LLZ1} for more details.

We would like to briefly explain the technical details of the proof.
The proof of the combinatorial cut-and-join formula is based on the
Burnside formula and various simple results in symmetric functions.
See \cite{Zho1}, \cite{LLZ0} and \cite{LLZ1}.

The proof of the geometric cut-and-join formula used the functorial
localization formula in \cite{LLY1} and \cite{LLY2}. The virtual
version of this formula was proved first applied to the virtual
fundamental cycles in the computations of Gromov-Witten invariants
in \cite{LLY2}.

As remarked in previous sections the functorial localization formula
is very effective and useful because we can use it to push
computations on complicated moduli space to simpler moduli space.
The moduli spaces used by mathematicians are usually the correct but
complicated moduli spaces like the moduli spaces of stable maps,
while the moduli spaces used by physicists are usually the simple
but the wrong ones like the projective spaces. This functorial
localization formula has been used successfully in the proof of the
mirror formula \cite{LLY1}, \cite{LLY2}, the proof of the Hori-Vafa
formula \cite{LLLY}, and the easy proof of the ELSV formula
\cite{LLZ1}. Our first proof of the Mari\~{n}o-Vafa formula also
used this formula in a crucial way.

More precisely, let ${\overline{\cal M}}_g(\mathbf{P}^1, \mu)$
denote the moduli space of
 relative stable maps from a genus $g$ curve to $\mathbf{P}^1$
 with fixed ramification type $\mu$ at $\infty$, where $\mu$ is a
 fixed partition. We apply the functorial localization formula to the divisor morphism
from the relative stable map moduli space to the projective space,

$$\mathrm{Br}:\ \ \overline{{\cal M}}_g(\mathbf{P}^1, \mu)\rightarrow
\mathbf{P}^r,$$ where $r$ denotes the dimension of $ \overline{{\cal
M}}_g(\mathbf{P}^1, \mu)$. This is similar to the set-up of mirror
principle, only with a different linearized moduli space, but in
both cases the target spaces are projective spaces.

We found that the fixed points of the target $\mathbf{P}^r$
precisely labels the cut-and-join operations of the triple Hodge
integrals. Functorial localization reduces the problem to the study
of polynomials in the equivariant cohomology group of
$\mathbf{P}^r$. We were able to squeeze out a system of linear
equations which implies the cut-and-join equation. Actually we
derived a stronger relation than the cut-and-join equation, while
the cut-and-join equation we need for the Mari\~{n}o-Vafa formula is
only the very first of such kind of relations. See \cite{LLZ1} for
higher order cut-and-join equations.

As was known in infinite Lie algebra theory, the cut-and-join
operator is closely related to and more fundamental than the
Virasoro algebras in some sense.

Recently there have appeared two different approaches to the
Mari\~{n}o-Vafa formula. The first one is a direct derivation of the
convolution formula which was discovered during our proof of the two
partition analogue of the formula \cite{LLZ2}. See \cite{CLiu} for
the details of the derivation in this case. The second is by
Okounkov-Pandhripande \cite{Oko-Pan}, they gave a different approach
by using the ELSV formula as initial value, and as well as the
$\lambda_g$ conjecture and other recursion relations from
localization on the moduli spaces of stable maps to $\mathbf{P}^1$.

\section{Two Partition Formula}

The two partition analogue of the Mari\~{n}o-Vafa formula naturally
arises from the localization computations of the Gromov-Witten
invariants of the open toric Calabi-Yau manifolds, as explained in
\cite{Zho3}.

To state the formula we let $\mu^+,  \mu^-$ be any two partitions.
Introduce the Hodge integrals involving these two partitions:

$$ G_{\mu^+, \mu^-}(\lambda; \tau) =B(\tau;\mu^+,\mu^-) \cdot\sum_{g \geq 0}
\lambda^{2g-2}A_g(\tau;\mu^+,\mu^-)$$where

$$A_g(\tau;\mu^+,\mu^-)=\int_{\Mbar_{g, l(\mu^+)+l(\mu^-)}}
\frac{\Lambda_{g}^{\vee}(1)\Lambda^{\vee}_{g}(\tau)\Lambda_{g}^{\vee}(-\tau
- 1)} {\prod_{i=1}^{l(\mu^+)}  \left(1 -\mu_i^+ \psi_i\right)
\prod_{j=1}^{l(\mu^-)}{\tau}\left(\tau -\mu_i^-
\psi_{j+l(\mu^+)}\right)}
$$and

$$B(\tau;\mu^+,\mu^-)=-\frac{(\sqrt{-1}\lambda)^{l(\mu^+)+l(\mu^-)}}{|\Aut({\mu^+})|
|\Aut({\mu^-})|} \left[\tau(\tau+1)\right]^{l(\mu^+)+l(\mu^-)-1}
\cdot$$

$$\prod_{i=1}^{l(\mu^+)} \frac{\prod_{a=1}^{\mu^+_i-1}\left(
\mu^+_i\tau + a \right)}{(\mu_i^+-1)!}\cdot\prod_{i=1}^{l(\mu^-)}
\frac{\prod_{a=1}^{\mu^-_i-1} \left( \mu_i^- \frac{1}{\tau} + a
\right)}{(\mu_i^--1)!}.$$ These complicated expressions naturally
arise in open string theory, as well as in the localization
computations of the Gromov-Witten invariants on open toric
Calabi-Yau manifolds.

We introduce two generating series, first on the geometry side,

$$ G^\bullet(\lambda;p^+,p^-;\tau)=\exp \left( \sum_{(\mu^+,
\mu^-) \in \cP^2} G_{\mu^+, \mu^-}(\lambda,
\tau)p^+_{\mu^+}p^-_{\mu^-}\right) ,$$ where $\cP^2$ denotes the set
of pairs of partitions and $p^\pm_{\mu^\pm}$ are two sets of formal
variables associated to the two partitions as in the last section.

On the representation side, we introduce

$$R^\bullet(\lambda;p^+,p^-;\tau)  = \sum_{|\nu^\pm|=|\mu^\pm| \geq
0} \frac{\chi_{\nu^+}(C(\mu^+))}{z_{\mu^+}}
\frac{\chi_{\nu^-}(C(\mu^-))}{z_{\mu^-}} \cdot
e^{\sqrt{-1}(\kappa_{\nu^+}\tau+ \kappa_{\nu^-}\tau^{-1})\lambda/2}
\cW_{\nu^+, \nu^-} p^+_{\mu^+}p^-_{\mu^-}.$$ Here

$$\cW_{\mu, \nu} = q^{l(\nu)/2} \cW_{\mu} \cdot s_{\nu}({\cal
E}_{\mu}(t))$$

 $$ =(-1)^{|\mu|+|\nu|}
q^{\frac{\kappa_{\mu}+\kappa_{\nu}+|\mu|+|\nu|}{2}} \sum_{\rho}
q^{-|\rho|} s_{\mu/\rho}(1, q, \dots)s_{\nu/\rho}(1, q, \dots)$$ in
terms of the skew Schur functions $s_\mu$ \cite{Mac}. They appear
naturally in the Chern-Simons invariant of the Hopf link.

\begin{theorem}{We have the identity:}

$$G^\bullet(\lambda;p^+,p^-;\tau)=R^\bullet(\lambda;p^+,p^-;\tau).$$

\end{theorem}

The idea of the proof is similar to that of the proof of the
Mari\~no-Vafa formula. We prove that both sides of the above
identity satisfy the same cut-and-join equation of the following
type:

$$\frac{\partial}{\partial \tau} { H}^{\bullet} =
\frac{1}{2}(CJ)^+H^{\bullet} - \frac{1}{2\tau^2}(CJ)^-H^{\bullet},$$
where $(CJ)^\pm$ denote the cut-and-join operator, the differential
operator with respect to the two set of variables $p^\pm$. We then
prove that they have the same initial value at $\tau=-1$:
$$G^{\bullet}(\lambda;p^+,p^-;-1) = R^{\bullet}(\lambda;p^+, p^-;-1),$$
which is again given by the Ooguri-Vafa formula \cite{LLZ2},
\cite{Zho3}.

The cut-and-join equation can be written in a linear matrix form,
and such equation follows from the convolution formula of the form

$$K_{\mu^+,\mu^-}^\bullet(\lambda)=\sum_{|\nu^\pm|=\mu^\pm}G^\bullet_{\mu^+,\mu^-}(\lambda;\tau)
z_{\nu^+}\Phi^\bullet_{\nu^+,\mu^+}(-\sqrt{-1}\lambda\tau)z_{\nu^-}\Phi^\bullet_{\nu^-,\mu^-}(\frac{-\sqrt{-1}}{\tau}\lambda)$$
where $\Phi^\bullet$ denotes the generating series of double Hurwitz
numbers, and $K_{\mu^+,\mu^-}$ is the generating series of certain
integrals on the moduli spaces of relative stable maps. For more
details see \cite{LLZ2}.

 This convolution formula arises naturally from localization
computations on the moduli spaces of relative stable maps to
$\mathbf{P}^1\times \mathbf{P}^1$ with the point $(\infty, \infty)$
blown up. So it reflects the geometric structure of the moduli
spaces. Such convolution type formula was actually discovered during
our search for a proof of this formula, both on the geometric and
the combinatorial side, see \cite{LLZ2} for the detailed derivations
of the convolution formulas in both geometry and combinatorics.

The proof of the combinatorial side of the convolution formula is
again a direct computation. The proof of the geometric side for the
convolution equation is to reorganize the generating series from
localization contributions on the moduli spaces of relative stable
maps into $\mathbf{P}^1\times \mathbf{P}^1$ with the point $(\infty,
\infty)$ blown up, in terms of the double Hurwitz numbers. It
involves careful analysis and computations.

\section{ Theory of Topological Vertex}

When we worked on the Mari\~{n}o-Vafa formula and its
generalizations, we were simply trying to generalize the method and
the formula to involve more partitions, but it turned out that in
the three partition case, we naturally met the theory of topological
vertex. Topological vertex was first introduced in string theory by
Vafa et al in \cite{AKMV}, it can be deduced from a three partition
analogue of the Mari\~{n}o-Vafa formula in a highly nontrivial way.
From this we were able to give a rigorous mathematical foundation
for the physical theory. Topological vertex is a high point of the
theory of string duality as developed by Vafa and his group for the
past several years, starting from Witten's conjectural duality
between Chern-Simons and open string theory. It gives the most
powerful and effective way to compute the Gromov-Witten invariants
for all open toric Calabi-Yau manifolds. In physics it is rare to
have two theories agree up to all orders, topological vertex theory
gives a very significant example. In mathematics the theory of
topological vertex already has many interesting applications. Here
we only briefly sketch the rough idea for the three partition
analogue of the Mari\~no-Vafa formula. For its relation to the
theory of topological vertex, we refer the reader to \cite{LLLZ} for
the details.

Given any three partitions $\overrightarrow{\mu}=\{\up{\mu}\}$, the
cut-and-join equation in this case, for both the geometry and
representation sides, has the form:

 $$\frac{\pa}{\pa\tau}F^\bullet(\lambda;\tau;\textbf{p}) =(CJ)^1 F^\bullet(\lambda;\tau;\textbf{p})
 +\frac{1}{\tau^2}(CJ)^2 F^\bullet(\lambda;\tau;\textbf{p})
 +\frac{1}{(\tau+1)^2}(CJ)^3 F^\bullet(\lambda;\tau;\textbf{p}).
$$
The cut-and-join operators $(CJ)^1$, $(CJ)^2$ and $(CJ)^3$ are with
respect to the three partitions. More precisely they correspond to
the differential operators with respect to the three groups of
infinite numbers of variables $\textbf{p}=\{p^1,p^2,p^3\}$.

 The initial value for this differential equation is taken at $\tau =1$,
 which is then reduced to the formulas of two partition case.
The combinatorial, or the Chern-Simons invariant side is given by
$\cW_{\overrightarrow{\mu}}=\cW_{\up{\mu}}$ which is a combination
of the $\cW_{\mu, \nu}$ as in the two partition case. See
\cite{LLLZ} for its explicit expression.

On the geometry side,
$$G^\bullet(\lambda;\tau;\textbf{p})=\texttt{exp}(G(\lambda;\tau;\textbf{p}))$$
is the non-connected version of the generating series of the triple
Hodge integral. More precisely,

$$G(\lambda;\tau;\textbf{p})=\sum_{\overrightarrow{\mu}}[\sum_{g=0}^\infty
\lambda^{2g-2+l(\overrightarrow{\mu})}G_{g,\overrightarrow{\mu}}(\tau)]p^1_{\mu^1}p^2_{\mu^2}p^3_{\mu_3}$$
where $l(\overrightarrow{\mu})=l(\mu^1)+l(\mu^2)+l(\mu^3)$ and
$G_{g,{\overrightarrow{\mu}}}(\tau)$ denotes the Hodge integrals of
the following form,

$$A(\tau)\,\int_{{\overline{\cal M}_{g,l_1+l_2+l_3}}}
\frac{\Lambda_g^\vee(1)\Lambda_g^\vee(\tau)\Lambda_g^\vee(-\tau-1)}
{\prod_{j=1}^{l_1}(1-\mu^1_j\psi_j)
 \prod_{j=1}^{l_2}\tau(\tau-\mu^2_j\psi_{l_1+j})}\cdot
 \frac{(\tau(\tau+1))^{l_1+l_2+l_3-1}}{\prod_{j=1}^{l_3}(\tau+1)(\tau+1+\mu^3_j\psi_{l_1+l_2+j})},$$
where

$$A(\tau)=\frac{-(\sqrt{-1}\lam)^{l_1+l_2+l_3}}{\ammm}\prod_{j=1}^{l_1}
  \frac{\prod_{a=1}^{\mu^1_j-1}(\tau\mu_j^1
  +a)}{(\mu_j^1-1)!}\cdot$$

  $$ \prod_{j=1}^{l_2}
  \frac{\prod_{a=1}^{\mu^1_j-1}((-1-1/\tau)\mu_j^2+a)}{(\mu_j^2-1)!}\prod_{j=1}^{l_3}
\frac{\prod_{a=1}^{\mu^1_j-1}(-\mu_j^3/(\tau+1)+a)}{(\mu_j^3-1)!}$$

In the above expression, $l_i=l(\mu^i)$, $i=1,2,3$. Despite of its
complicated coefficients, these triple integrals naturally arise
from localizations on the moduli spaces of relative stable maps into
the blow-up of  $\mathbf{P}^1\times \mathbf{P}^1\times \mathbf{P}^1$
along certain divisors. It also naturally appears in open string
theory computations \cite{AKMV}. See \cite{LLLZ} for more details.

One of our results in \cite{LLLZ} states that
$G^\bullet(\lambda;\tau;\textbf{p})$ has a combinatorial expression
$R^\bullet(\lambda;\tau;\textbf{p})$ in terms of the Chern-Simons
knot invariants $W_{\overrightarrow{\mu}}$, which is a closed
combinatorial expression. More precisely it is given by

$$R^\bullet(\lambda;\tau;\textbf{p})=\sum_{\overrightarrow{\mu}}[\sum_{|\nu^i|=|\mu^i|}
\prod_{i=1}^3\frac{\chi_{\nu^i}(\mu^i)}{z_{\mu^i}}
q^{\frac{1}{2}(\sum^3_{i=1}\kappa_{\nu^i}\frac{w_{i+1}}{w_i})}W_{\overrightarrow{\nu}}(q)]p^1_{\mu^1}p^2_{\mu^2}p^3_{\mu^3}.$$
Here $w_4=w_1$ and $w_3=-w_1-w_2$ and $\tau=\frac{w_2}{w_1}$. Due to
the complicated combinatorics in the initial values, the
combinatorial expression $W_{\overrightarrow{\mu}}$ we obtained is
different from the expression $\cW_{\overrightarrow{\mu}}$ obtained
by Vafa et al. Actually our expression is even simpler than theirs
in some sense. The expression we obtained is more convenient for
mathematical applications such as the proof of the Gopakumar-Vafa
conjecture for open toric Calabi-Yau manifolds, see \cite{Pan}.

\begin{theorem} We have the equality:

$$G^\bullet(\lambda;\tau;\textbf{p})=R^\bullet(\lambda;\tau;\textbf{p}).$$
\end{theorem}

The key point to prove the above theorem is still the proof of
convolution formulas for both sides which imply the cut-and-join
equation. The proof of the convolution formula for
$G^\bullet(\lambda;\tau;\textbf{p})$ is much more complicated than
the one and two partition cases. See \cite{LLLZ} for details.

The most useful property of topological vertex is its gluing
property induced by the orthogonal relations of the characters of
the symmetric group. This is very close to the situation of two
dimensional gauge theory. In fact string theorists consider
topological vertex as a kind of lattice theory on Calabi-Yau
manifolds. By using the gluing formula we can easily obtain closed
formulas for generating series of Gromov-Witten invariants of all
genera and all degrees, open or closed, for all open toric
Calabi-Yau manifolds, in terms of the Chern-Simons knot invariants.
Such formulas are always given by finite sum of products of those
Chern-Simons type invariants $\cal W_{\mu,\nu}$'s. The magic of
topological vertex is that, by simply looking at the moment map
graph of the toric surfaces in the open toric Calabi-Yau, we can
immediately write down the closed formula for the generating series
for all genera and all degree Gromov-Witten invariants, or more
precisely the Euler numbers of certain bundles on the moduli space
of stable maps.

Here we only give one example to describe the topological vertex
formula for the generating series of the all degree and all genera
Gromov-Witten invariants for the open toric Calabi-Yau 3-folds. We
write down the explicit close formula of the generating series of
the  Gromov-Witten invariants in this case.

\medskip

\paragraph{\bf Example:} Consider the toric Calabi-Yau manifold
 which is $O(-3)\longrightarrow \mathbf{P}^2$.
 In this case the formula for the generating series of all degrees and all genera
 Gromov-Witten invariants is given by

$$\mathrm{exp}\,( \sum_{g=0}^\infty \lambda^{2g-2}F_g(t))= \sum_{\nu_1, \nu_2,\nu_3} \cW_{\nu_1,\nu_2}\cW_{\nu_2,\nu_3}
\cW_{\nu_3, \nu_1}
(-1)^{\sum_{j=1}^3|\nu_j|}q^{\frac{1}{2}\sum_{i=1}^3
\kappa_{\nu_i}}\, e^{t(\sum_{j=1}^3|\nu_j|)}$$ where
$q=e^{\sqrt{-1}\lambda}$. The precise definition of $F_g(t)$ will be
given in the next section.

For general open toric Calabi-Yau manifolds, the expressions are
just similar. They are all given by finite and closed formulas,
which are easily read out from the moment map graphs associated to
the toric surfaces, with the topological vertex associated to each
vertex of the graph.

 In \cite{AKMV} Vafa and his group first developed the theory of topological vertex by using
string duality between Chern-Simons and Calabi-Yau, which is a
physical theory. In \cite{LLLZ} we established the mathematical
theory of the topological vertex, and derived various mathematical
corollaries, including the relation of the Gromov-Witten invariants
to the equivariant index theory as motivated by the Nekrasov
conjecture in string duality \cite{LLZ}. During the development of
the mathematical theory of topological vertex we also introduced
formal Calabi-Yau manifolds, see \cite{LLLZ} for details.

\section{ Gopakumar-Vafa Conjecture and Indices of
Elliptic Operators }

Let $N_{g, d}$ denote the so-called Gromov-Witten invariant of genus
$g$ and degree $d$ of an open toric Calabi-Yau 3-fold. $N_{g, d}$ is
defined to be the Euler number of the obstruction bundle on the
moduli space of stable maps of degree $d \in H_2(S, \mathbb{Z})$
from genus $g$ curve into the surface base $S$. The open toric
Calabi-Yau manifold associated to the toric surface $S$ is the total
space of the canonical line bundle $K_S$ on $S$. More precisely

$$N_{g, d} =\int_{[\Mbar_g(S, d)]^v} e(V_{g,d})$$
with $V_{g,d} =R^1\pi_*u^*K_S$ a vector bundle on the moduli space
induced by the canonical bundle $K_S$. Here $\pi: \ U\rightarrow
\Mbar_g(S, d)$ denotes the universal curve and $u$ can be considered
as the evaluation or universal map. Let us write

$$F_g(t) =\sum_{d\geq 0}N_{g, d}\,e^{-d\cdot t}.$$

The Gopakumar-Vafa conjecture is stated as follows:
\begin{conj}
 {\em There exists an expression:
$$\sum_{g=0}^\infty \lambda^{2g-2}F_g(t)=\sum_{k=1}^\infty\sum_{g,
d\geq
0}n_d^{g}\,\frac{1}{d}\,(2\,{\mathrm{sin}}\,\frac{d\lambda}{2})^{2g-2}e^{-kd\cdot
t},$$ such that $n_d^g$ are integers, called instanton numbers.}

\end
{conj}
Motivated by the Nekrasov duality conjecture between the four
dimensional gauge theory and string theory, we are able to interpret
the above integers $n_d^g$ as equivariant indices of certain
elliptic operators on the moduli spaces of anti-self-dual
connections \cite{LLZ}:

\begin{theorem} {For certain interesting cases, these $n_d^g$'s can be written
as equivariant indices on the moduli spaces of anti-self-dual
connections on $\mathbb{C}^2$.}
\end{theorem}

For more precise statement, we refer the reader to \cite{LLZ0}. The
interesting cases include open toric Calabi-Yau manifolds when $S$
is Hirzebruch surface. The proof of this theorem is to compare fixed
point formula expressions for equivariant indices of certain
elliptic operators on the moduli spaces of anti-self-dual
connections with the combinatorial expressions of the generating
series of the Gromov-Witten invariants on the moduli spaces of
stable maps. They both can be expressed in terms of Young diagrams
of partitions. We find that they agree up to certain highly
non-trivial "mirror transformation", a complicated variable change.
This result is not only interesting for the index formula
interpretation of the instanton numbers, but also for the fact that
it gives the first complete examples that the Gopakumar-Vafa
conjecture holds for all genera and all degrees.

Recently P. Peng \cite{Pan} has given the proof of the
Gopakumar-Vafa conjecture for all open toric Calabi-Yau 3-folds by
using our Chern-Simons expressions from the topological vertex. His
method is to explore the property of the Chern-Simons expression in
great detail with some clever observation about the form of the
combinatorial expressions. On the other hand, Kim in \cite{Kim} has
derived some remarkable recursion formulas for Hodge integrals of
all genera and any number of marked points, involving one
$\lambda$-classes. His method is to add marked points in the moduli
spaces and then follow the localization argument we used to prove
the Mari\~{n}o-Vafa formula.

\section{Two Proofs of the ELSV Formula}
\label{sec:ELSVpf}

In this section we describe two proofs of the ELSV formula, one is
by direct localization and cut-and-join equation following our proof
of the Mari\~no-Vafa formula, another one is to derive it from the
Mari\~no-Vafa formula through a scaling limit.

Given a partition $\mu$ of length $l(\mu)$, denote by $H_{g_, \mu}$
the Hurwitz numbers of almost simple Hurwitz covers of $\bP^1$ of
ramification type $\mu$ by  connected genus $g$ Riemann surfaces.
The ELSV formula \cite{ELSV, Gra-Vak} states:
$$
H_{g, \mu} =(2g-2+|\mu|+l(\mu))! I_{g,\mu}
$$
where
$$
I_{g,\mu}=\frac{1}{|\Aut (\mu)|} \prod_{i=1}^{l(\mu)}
\frac{\mu_i^{\mu_i}}{\mu_i!} \int_{\Mbar_{g, l(\mu)}}
\frac{\Lambda_g^{\vee}(1)}{\prod_{i=1}^{l(\mu)} (1 - \mu_i \psi_i)}.
$$

Define generating functions
\begin{eqnarray*}
\Phi_{\mu}(\lambda)&=& \sum_{g \geq 0} H_{g, \mu}
\frac{\lambda^{2g-2+|\mu|+l(\mu)}}{(2g-2 + |\mu| + l(\mu))!}, \\
\Phi(\lambda; p) &=& \sum_{|\mu|\geq 1} \Phi_{\mu}(\lambda) p_{\mu}, \\
\Psi_{\mu}(\lambda) &=&\sum_{g \geq 0}I_{g,\mu}
\lambda^{2g-2+|\mu|+l(\mu)},\\
\Psi(\lambda; p) &=& \sum_{|\mu|\geq 1} \Psi_{\mu}(\lambda)p_{\mu}.
\end{eqnarray*}
In terms of generating functions,  the ELSV formula reads
\begin{equation*}\label{eqn:ELSV}
\Psi(\lambda;p)=\Phi(\lambda;p).
\end{equation*}

It was known that $\Phi(\lambda;p)$ satisfies the following
cut-and-join equation:
\begin{equation*}\label{eqn:ELSVcj}
\frac{\partial \Theta}{\partial \lambda} = \frac{1}{2} \sum_{i,
j\geq 1} \left(ijp_{i+j}\frac{\partial^2\Theta}{\partial p_i\partial
p_j} + ijp_{i+j}\frac{\partial \Theta}{\partial p_i}\frac{\partial
\Theta}{\partial p_j} + (i+j)p_ip_j\frac{\partial \Theta}{\partial
p_{i+j}}\right).
\end{equation*}
Later this equation was reproved by sum formula of symplectic
Gromov-Witten invariants \cite{Li-Zha-Zhe}.

The calculations in Section 7 and Appendix A of \cite{LLZ} shows
that
\begin{equation*}\label{eqn:HI}
\tilde{H}_{g,\mu}= (2g-2+|\mu|+l(\mu))! I_{g,\mu}
\end{equation*}
\begin{equation*}\label{eqn:HIone}
\begin{split}
& \tilde{H}_{g,\mu}=(2g-3+|\mu|+l(\mu))!\left(
\sum_{\nu\in J(\mu)} I_{g,\nu}+\sum_{\nu\in C(\mu)} I_2(\nu) I_{g-1,\nu}\right.\\
&\makebox[3cm]{ } +\left.\sum_{g_1+g_2=g}\sum_{\nu^1\cup \nu^2\in
C(\mu)} I_3(\nu^1,\nu^2)I_{g_1,\nu_1} I_{g_2,\nu_2}\right)
\end{split}
\end{equation*}
where
$$
\tilde{H}_{g,\mu}=\int_{[\Mbar_{g,0}(\P^1,\mu)]^{\mathrm{vir} }
}\mathrm{Br}^* H^r
$$
is some relative Gromov-Witten invariant of $(\P^1,\infty)$, and
$C(\mu), J(\mu), I_1, I_2,I_3$ are defined as in \cite{Li-Zha-Zhe}.
So we  have
\begin{eqnarray*}
 &&(2g-2+|\mu|+l(\mu)) I_{g,\mu}\\
 &=&
\sum_{\nu\in J(\mu)} I_{g,\nu}+\sum_{\nu\in C(\mu)} I_2(\nu)
I_{g-1,\nu} +\sum_{g_1+g_2=g}\sum_{\nu^1\cup \nu^2\in C(\mu)}
I_3(\nu^1,\nu^2)I_{g_1,\nu_1} I_{g_2,\nu_2},
\end{eqnarray*}
which is equivalent to the statement that the generating function
$\Psi(\lambda;p)$ of $I_{g,\mu}$ also satisfies the cut-and-join
equation.

Any solution $\Theta(\lambda;p)$ to the cut-and-join equation
(\ref{eqn:ELSVcj}) is uniquely determined by its initial value
$\Theta(0;p)$, so it remains to show that $\Psi(0;p)=\Phi(0;p)$.
Note that $2g-2+|\mu|+l(\mu)=0$ if and only if $g=0$ and $\mu=(1)$,
so
$$
\Psi(0;p)=H_{0,(1)}p_1,\ \ \Phi(0;p)=I_{0,(1)}p_1.
$$
It is easy to see that $H_{0,(1)}=I_{0,(1)}=1$, so
$$
\Psi(0;p)=\Phi(0;p).
$$

One can see geometrically that the relative Gromov-Witten invariant
$\tilde{H}_{g,\mu}$ is equal to the Hurwitz number $H_{g,\mu}$. This
together with (\ref{eqn:HI}) gives a proof of the ELSV formula
presented in \cite[Section 7]{LLZ} in the spirit of \cite{Gra-Vak}.
Note that $\tilde{H}_{g,\mu}=H_{g,\mu}$  is not used in the proof
described above.

On the other hand we can deduce the ELSV formula as the limit of the
Mari\~{n}o-Vafa formula. By the Burnside formula, one easily gets
the following expression (see e.g. \cite{LLZ2}):
\begin{eqnarray*}
\Phi(\lambda;p) &=&\log\left(\sum_\mu\left(\sum_{|\nu|=|\mu|}
\frac{\chi_{\nu}(\mu)}{z_{\mu}} e^{\kappa_{\nu}\lambda/2} \frac{\dim
R_{\nu}}{|\nu|!}\right) p_{\mu}.
\right)\\
& = & \sum_{ n \geq 1} \frac{(-1)^{n-1}}{n} \sum_{\mu}
\sum_{\cup_{i=1}^n \mu_i = \mu} \prod_{i=1}^n \sum_{|\nu_i|=|\mu_i|}
\frac{\chi_{\nu_i}(\mu_i)}{z_{\mu_i}} e^{\kappa_{\nu_i}\lambda/2}
\frac{\dim R_{\nu_i}}{|\nu_i|!} p_{\mu}.
\end{eqnarray*}
The ELSV formula reads
$$
\Psi(\lambda;p)=\Phi(\lambda;p)
$$
where the left hand side is a generating function of Hodge integrals
$I_{g,\mu}$, and the right hand side is a generating function of
representations of symmetric groups. So the ELSV formula and the
Mari\~no-Vafa formula are of the same type.

Actually, the ELSV formula can be obtained by taking a particular
limit of the Mari\~no-Vafa formula
$G(\lambda;\tau;p)=R(\lambda;\tau;p)$. More precisely, it is
straightforward to check that
\begin{eqnarray*}
&& \lim_{\tau\to 0}G(\lambda\tau;\frac{1}{\tau};
(\lambda\tau)p_1,(\lambda\tau)^2 p_2,\cdots)\\
&=&\sum_{|\mu|\neq 0}
\sum_{g=0}^\infty \sqrt{-1}^{2g-2+|\mu|+l(\mu)} I_{g,\mu}\lambda^{2g-2+|\mu|+l(\mu)}p_\mu \\
&=& \Psi(\sqrt{-1}\lambda;p)
\end{eqnarray*}
and
\begin{eqnarray*}
&& \lim_{\tau\to 0}R(\lambda\tau;\frac{1}{\tau};
(\lambda\tau)p_1,(\lambda\tau)^2 p_2,\cdots)\\
&=&\log\left(\sum_{\mu}\left(\sum_{|\nu|=|\mu|}\frac{\chi_{\nu}(C(\mu))}{z_{\mu}}
e^{\sqrt{-1}\kappa_{\nu}\lambda/2}\lim_{t\to 0}
 (t^{|\nu|}V_{\nu}(t))\right) p_\mu\right)\\
&=&\log\left(\sum_{\mu}\left(\sum_{|\nu|=|\mu|}\frac{\chi_{\nu}(C(\mu))}{z_{\mu}}
e^{\sqrt{-1}\kappa_{\nu}\lambda/2} \frac{1}{\prod_{x\in\nu}h(x)}\right) p_\mu \right)\\
&=&\Phi(\sqrt{-1}\lambda;p)
\end{eqnarray*}
where we have used
$$
\frac{1}{\prod_{x\in\nu}h(x)}=\frac{\dim R_\nu}{|\nu|!}.
$$

In this limit, the cut-and-join equation of $G(\lambda;\tau;p)$ and
$R(\lambda;\tau;p)$ reduces to the cut-and-join equation of
$\Psi(\lambda;p)$ and $\Phi(\lambda;p)$, respectively.

\section{A Localization Proof of the Witten Conjecture}

The Witten conjecture for moduli spaces states that the generating
series $F$ of the integrals of the $\psi$ classes for all genera and
any number of marked points satisfies the KdV equations and the
Virasoro constraint. For example the Virasoro constraint states that
$F$ satisfies
$$L_n\cdot F=0, \ n\geq -1
$$ where $L_n$ denote certain Virasoro operators to be given later.

Witten conjecture was first proved by Kontsevich using combinatorial
model of the moduli space and matrix model, with later approaches by
Okounkov-Pandhripande using ELSV formula and combinatorics,  by
Mirzakhani using Weil-Petersson volumes on moduli spaces of bordered
Riemann surfaces.

I will present a much simpler proof by using functorial localization
and asymptotics. This was done jointly with Y.-S. Kim in \cite{KL}.
This is also motivated by methods in proving conjectures from string
duality. It should have more applications.

The basic idea of our proof is to directly prove the following
recursion formula which, as derived in physics by Dijkgraaf,
Verlinde and Verlinde by using quantum field theory, implies the
Virasoro and the KdV equation for the generating series $F$ of the
integrals of the $\psi$ classes:

\begin{theorem}
{ We have identity}
$$\langle\tilde{\sigma}_{n}\prod_{k\in
S}\tilde{\sigma}_{k}\rangle_{g}=
        \sum_{k\in S}(2k+1)\langle\tilde{\sigma}_{n+k-1}\prod_{l\neq
        k}\tilde{\sigma}_{l}\rangle_{g}+\frac{1}{2}\sum_{a+b=n-2}\langle\tilde{\sigma}_{a}\tilde{\sigma}_{b}\prod_{l\neq
a,b}\tilde{\sigma}_{l}\rangle_{g-1}$$
$$+\frac{1}{2}\sum_{\stackrel{S=X\cup
Y,}{\stackrel{a+b=n-2,}{g_{1}+g_{2}=g}}}
        \langle\tilde{\sigma}_{a}\prod_{k\in X}\tilde{\sigma}_{k}\rangle_{g_{1}}
        \langle\tilde{\sigma}_{b}\prod_{l\in
        Y}\tilde{\sigma}_{l}\rangle_{g_{2}}.$$

\end{theorem}

Here $\tilde{\sigma}_{n}=(2n+1)!!\psi^{n}$ and
$$ \langle\prod_{j=1}^n\tilde{\sigma}_{k_j}\rangle_{g}= \int_{\overline{\mathcal{M}}_{g,n}}
            \prod_{j=1}^n\tilde{\sigma}_{k_{j}}.$$
The notation $S=\{k_1, \cdots, k_n\}=X\cup Y$.

 To prove the above recursion relation, similar to the proof of the Mari\~no-Vafa formula,
  we first apply the functorial
localization to the natural branch map from moduli space of relative
stable maps $\overline{\mathcal{M}}_{g}(\mathbf{P}^{1},\mu)$ to
projective space $\mathbf{P}^r$ where $r=2g-2+\vert\mu\vert+l(\mu)$
is the dimension of the moduli.

As discussed in last section we easily get the cut-and-join equation
for one Hodge integral
$$I_{g, \mu}=\frac{1}{\vert\text{Aut }\mu\vert}\prod_{i=1}^{n}\frac{\mu_{i}^{\mu_{i}}}{\mu_{i}!}
                \int_{\overline{\mathcal{M}}_{g,n}}\frac{\Lambda_{g}^{\vee}(1)}{\prod(1-\mu_{i}\psi_{i})}.$$
The equation we get has the form as discussed in last section, it is
trivial corollary of the fact that the push-forward of $1$ in
equivariant cohomology by a map between equal dimension manifolds is
a constant:
\begin{eqnarray*}
 &&(2g-2+|\mu|+l(\mu)) I_{g,\mu}\\
 &=&
\sum_{\nu\in J(\mu)} I_{g,\nu}+\sum_{\nu\in C(\mu)} I_2(\nu)
I_{g-1,\nu} +\sum_{g_1+g_2=g}\sum_{\nu^1\cup \nu^2\in C(\mu)}
I_3(\nu^1,\nu^2)I_{g_1,\nu_1} I_{g_2,\nu_2}.
\end{eqnarray*}
 Note that more general formulas of such type was first
found and proved by Kim in \cite{Kim}.

Write $\mu_i=Nx_i$. Let $N$ go to infinity and expand in $x_i$, we
get:
$$\sum_{i=1}^{n}\Big[\frac{(2k_{i}+1)!!}{2^{k_{i}+1}k_{i}!}x_{i}^{k_{i}}
            \prod_{j\neq i}\frac{x_{j}^{k_{j}-\frac{1}{2}}}{\sqrt{2\pi}}\int_{\overline{\mathcal{M}}_{g,n}}
            \prod\psi_{j}^{k_{j}}-\sum_{j\neq
i}\frac{(x_{i}+x_{j})^{k_i+k_j-\frac{1}{2}}}{\sqrt{2\pi}}\prod_{l\neq
i,j}
            \frac{x_{l}^{k_{l}-\frac{1}{2}}}{\sqrt{2\pi}}\int_{\overline{\mathcal{M}}_{g,n-1}}
            \psi^{k_i+k_j-1}\prod\psi_{l}^{k_{l}}$$
            $$-\frac{1}{2}\sum_{k+l=k_{i}-2}\frac{(2k+1)!!(2l+1)!!}{2^{k_i}\, k_i!}
            x_{i}^{k_i}\prod_{j\neq
            i}\frac{x_{j}^{k_{j}-\frac{1}{2}}}{\sqrt{2\pi}}\Big[\int_{\overline{\mathcal{M}}_{g-1,n+1}}\psi_{1}^{k}\psi_{2}^{l}\prod\psi_{j}^{k_{j}}$$
           $$ +\sum_{\stackrel{g_{1}+g_{2}=g,}{
\nu_{1}\cup\nu_{2}=\nu}}\int_{\overline{\mathcal{M}}_{g_{1},n_{1}}}
            \psi_{1}^{k}\prod\psi_{j}^{k_{j}}\int_{\overline{\mathcal{M}}_{g_{2},n_{2}}}\psi_{1}^{l}
            \prod\psi_{j}^{k_{j}}\Big]\Big]=0.$$
Performing Laplace transforms on the $x_i$'s, we get the recursion
formula in the above theorem which implies both the KdV equations
and the Virasoro constraints. For example the Virasoro constraints
states that the generating series

 $$\tau(\tilde{t})=\text{exp}\sum_{g=0}^{\infty}\langle\text{exp}\sum_{n}\tilde{t}_{n}\tilde{\sigma}_{n}\rangle_{g}$$
satisfies the equations:
    $$L_{n}\cdot\tau=0,\qquad (n\geq -1)$$
    where $L_{n}$ denote the Virasoro differential operators
    \begin{align*}
        L_{-1} &= -\frac{1}{2}\frac{\partial}{\partial\tilde{t}_{0}}+
            \sum_{k=1}^{\infty}(k+\frac{1}{2})\tilde{t}_{k}\frac{\partial}{\partial\tilde{t}_{k-1}}
            +\frac{1}{4}\tilde{t}^{2}_{0}\\
        L_{0} &= -\frac{1}{2}\frac{\partial}{\partial\tilde{t}_{1}}+
            \sum_{k=0}^{\infty}(k+\frac{1}{2})\tilde{t}_{k}\frac{\partial}{\partial\tilde{t}_{k}}+\frac{1}{16}\\
        L_{n} &= -\frac{1}{2}\frac{\partial}{\partial\tilde{t}_{n-1}}+
            \sum_{k=0}^{\infty}(k+\frac{1}{2})\tilde{t}_{k}\frac{\partial}{\partial\tilde{t}_{k+n}}
            +\frac{1}{4}\sum_{i=1}^{n}\frac{\partial^{2}}{\partial\tilde{t}_{i-1}\partial\tilde{t}_{n-i}}
    \end{align*}
We remark the same method can be used to derive very general
recursion formulas in Hodge integrals and general Gromov-Witten
invariants. We hope to report these results on a later occasion.
\section{Final Remarks}

We strongly believe that there is a more interesting and grand
duality picture between Chern-Simons invariants for three
dimensional manifolds and the Gromov-Witten invariants for open
toric Calabi-Yau manifolds. Our proofs of the Mari\~{n}o-Vafa
formula, and the setup of the mathematical foundation for
topological vertex theory and the results of others we have
discussed above all together have just opened a small window for a
more splendid picture. We can expect more exciting conjectures from
such duality to stimuate more developments in mathematics.

\end{document}